\documentclass[a4paper,11pt]{article}
\usepackage{jheppub} 
\usepackage{lineno}

\usepackage{braket}
\usepackage{bm}

\title{\boldmath The spectrum of perturbed $(3,10)$ minimal model}

\author{Andrei Katsevich}
\affiliation{Joseph Henry Laboratories, Princeton University, Princeton, NJ 08544, USA}

\emailAdd{akatsevich@princeton.edu}

\abstract{We study RG flows between non-unitary minimal models and massive quantum theories using Truncated Conformal Space Approach (TCSA). We consider the integrable non-unitary Yang-Lee model perturbed by $i\phi$ and the $D$-series version of $M(3,10)$ which is a product of two Yang-Lee models, perturbing the latter by relevant operators $\phi_{1,3}$ and $i\phi^+_{1,5}$. Utilizing the quasi-primary fields we find, TCSA is performed up to the level $N=15$ for $M(2,5)+i\phi$. The conjecture about the $M(3,10)$ perturbed by $\phi_{1,3}$ is stated: this theory flows to a massive phase; its spectrum contains a kink and two breathers, whose masses we find. Our TCSA results support the conjecture.}

\begin{document}
\maketitle
\flushbottom

\section{Introduction}
Conformal field theory (CFT) plays a significant role in investigating the universality classes of quantum field theories at critical points. The minimal models $M(p,q)$, defined for coprime positive integers $p$ and $q$, are an infinite class of two-dimensional CFTs that were defined by Belavin, Polyakov, and Zamolodchikov in the landmark paper \cite{Belavin:1984vu}. The Hilbert space of such theories is built from finitely many irreducible representations of the Virasoro algebra \cite{zbMATH03661547}.

Recently, the non-unitary minimal models ($|q-p|>1$) have been attracting more interest \cite{Xu:2022mmw,Xu:2023nke,Xu:2024baz,Nakayama:2022svf,Nakayama:2024msv,Klebanov:2022syt,Katsevich:2024jgq,Delouche:2023wsl,Lencses:2022ira,Lencses:2023evr,Lencses:2024wib}. The Kac table of non-unitary models contains fields with negative dimensions, and some of the structure constants entering the Operator Product Expansions (OPE) have imaginary values. The first representative is $M(2,5)$ which describes the class of universality of the Yang-Lee edge singularity of the zeros of the grand canonical partition function of the Ising model \cite{Yang:1952be,Lee:1952ig,Fisher:1978pf,Cardy:1985yy}. See also \cite{Cardy:2023lha} for a recent review.

Two-dimensional Integrable Quantum Field Theories (IQFTs) may be represented as Ultraviolet (UV) CFTs perturbed by some relevant operator \cite{Zamolodchikov:1989fp,Zamolodchikov:1989hfa}. The simplest example is the limit $T\rightarrow T_c$ in the Ising model with zero magnetic field $h=0$. This is an integrable theory of free Majorana fermions which might be considered as a unitary minimal model $M(3,4)$ perturbed by a spinless primary field $\phi_{1,3}\equiv\epsilon$.

Renormalization Group (RG) flows are a fundamental tool in theoretical physics, particularly in the study of QFTs and Statistical Mechanics \cite{Wilson:1973jj}. RG flow which begins at a UV CFT may end at an Infrared (IR) massive theory or another fixed point which is also conformal. The example of the first case is the non-unitary minimal model $M(2,2n+1)$ perturbed by primary field $\phi_{1,3}$. In particular, perturbed $M(2,5)+i\phi$ is an IQFT with spectrum consisting of one particle (breather) \cite{Cardy:1989fw}; it is called Scaling Lee-Yang Model (SLYM). The example of the second case is RG flow $M(3,4)+m\epsilon+ih\sigma\rightarrow M(2,5)$ \cite{Fonseca:2001dc,Xu:2022mmw}. 

Conformal Perturbation Theory (ConfPT) is used to study how a CFT behaves when it is slightly altered by a nearly marginal operator. It helps us understand the behavior of a system near its critical points and the effects of small deviations from conformality. Zamolodchikov used ConfPT to describe RG flow between unitary models \cite{Zamolodchikov:1987ti}: $M(p,p+1)+\phi_{1,3}\rightarrow M(p-1,p)$. 

Sometimes ConfPT is not working, and we need to utilize nonperturbative methods, for instance, the Truncated Conformal Space Approach (TCSA). It was originally applied for SLYM in \cite{Yurov:1989yu}. The main idea of this approach is to truncate Hilbert space up to a finite number of states. A modification of the TCSA is the Truncated Free Fermion Space Approach (TFFSA) \cite{Yurov:1991my,Fonseca:2001dc} which was created for perturbed $M(3,4)$ using the basis of massive free fermions. In paper \cite{Kausch:1996vq}, spectra of perturbed non-unitary minimal models $M(3,14)+\phi_{1,5}$, $M(3,16)+\phi_{1,5}$ were considered; also, model $M(3,10)+i\phi^+_{1,5}$ was discussed. A $D_6$ series version \cite{Cappelli:1986hf} of non-unitary $M(3,10)$ model is a tensor product of two Yang-Lee models $M(2,5)$ \cite{Kausch:1996vq,Quella:2006de,Ardonne_2011}. See also more recent publications about the use of TCSA for the RG flows between Yang-Lee singularities \cite{Xu:2022mmw,Lencses:2022ira,Lencses:2023evr,Lencses:2024wib}. The non-unitary RG flow $M(3.7)+i\phi_{1,3}+\phi_{1,5}\rightarrow M(3,5)$ \cite{Delouche:2023wsl} was investigated using Hamiltonian Truncation for QFTs with UV divergences \cite{EliasMiro:2022pua}.

There are some general statements about the existence of RG flows between minimal models. Recently, using the topological defect line anomaly matching \cite{Chang:2018iay}, Nakayama and Tanaka proposed a set of RG flows $M(kq+I,q)+\phi_{1,2k+1}\rightarrow M(kq-I,q)$ \cite{Nakayama:2024msv}. This includes the unitary family $k=1, I=1$ considered in \cite{Zamolodchikov:1987ti} by Zamolodchikov; the $k=1, I>1$ family studied in \cite{Lassig:1991an,Ahn:1992qi} by Lassig and Ahn; the $k=2$ family in \cite{Martins:1992ht,Martins:1992yk,Ravanini:1994pt,Dorey:2000zb} by Dorey et al. The first known example of $k=3$ flow $M(3,10)+\phi_{1,7}\rightarrow M(3,8)$ was suggested  in \cite{Fei:2014xta} and discussed further in \cite{Klebanov:2022syt,Katsevich:2024jgq}. Its generalization $M(q,3q+1)+\phi_{1,7}\rightarrow M(q,3q-1)$ was considered in \cite{Katsevich:2024jgq} from the GL point of view. Also, there is a half-integer $k=\frac{1}{2}$ family of RG flows: $M(\frac{q}{2}+\frac{I}{2},q)+\phi_{2,1}\rightarrow M(\frac{q}{2}-\frac{I}{2},q)$ \cite{Martins:1992ht,Martins:1992yk,Ravanini:1994pt,Dorey:2000zb}.

All the unitary RG flows satisfy Zamolodchikov's $c$-theorem \cite{Zamolodchikov:1986gt}. There is an extension of this theorem to non-unitary RG flows if the $\mathcal{PT}$ symmetry is preserved \cite{Castro-Alvaredo:2017udm}. In the extended theorem, the role of monotonically decreasing along the RG flow value plays an effective central charge $c_\text{eff}$, so $c^{\text{IR}}_\text{eff}\leq c^{\text{UV}}_\text{eff}$.

In this paper, we apply TCSA to SLYM and $M(3,10)$ perturbed by relevant fields $\phi_{1,3}$ and $i\phi^+_{1,5}$. In the landmark work \cite{Yurov:1989yu} TCSA was done for SLYM up to the level $N=5$. Finding quasi-primary fields in $M(2,5)$, we will reproduce classic results using TCSA up to the level $N=15$. Using the correspondence between the reduced sine-Gordon theory and minimal model $M(2,2n+1)$ perturbed by relevant field $\phi_{1,3}$, we will make the conjecture about the spectrum of $M(3,10)+\phi_{1,3}$: it consists of a kink with mass $m_{\text{kink}}$ and two breathers with masses $m_1=2\sin\frac{3\pi}{14}m_{\text{kink}}$ and $m_2=2\sin\frac{3\pi}{7}m_{\text{kink}}$. It will be checked numerically by TCSA. Also, using TCSA, we will check the well-known spectrum of $M(3,10)+i\phi^+_{1,5}$ \cite{Kausch:1996vq}: the model contains two breathers with equal mass.

The layout of the paper is organized as follows. In section~\ref{sec:review}, we briefly review the general information about minimal models. In section \ref{sec:predictions}, we consider theoretical results for the spectra of SLYM, $M(3,10)$ perturbed by $i\phi^+_{1,5}$ and make a conjecture about the spectrum of $M(3,10)$ perturbed by $\phi_{1,3}$. In section \ref{sec:TCSA}, we describe the TCSA algorithm and in section \ref{sec:Application}, we apply it to obtain the results.

\section{Review of the \texorpdfstring{$M(2,5)$}{M(2,5)} and \texorpdfstring{$M(3,10)$}{M(3,10)}}
\label{sec:review}
Kac table of minimal model $M(p,q)$ consists of $\frac{1}{2}(p-1)(q-1)$ primary fields $\phi_{m,n}$. Central charge of $M(p,q)$:
\begin{equation}
    c(p,q)=1-\frac{6(p-q)^2}{pq}\,.
\end{equation}

Holomorphic dimension of primary field $\phi_{m,n}$:
\begin{equation}
    h_{m,n}=\frac{(np-mq)^2-(p-q)^2}{4pq}=\frac{(np-mq)^2}{4pq}+\frac{c(p,q)-1}{24}\,,
\end{equation}
where $0<m<p$ and $0<n<q$. The primary fields are identified according to $\phi_{m,n}\equiv\phi_{p-m,q-n}$.

For non-unitary minimal models the effective central charge:
\begin{equation}
    c_\text{eff}(p,q)=c(p,q)-24h_\text{min}=1-\frac{6}{pq}\,,
\end{equation}
where $h_\text{min}$ is the lowest primary field dimension. For unitary models it is always $h_\text{min}=0$, so $c_\text{eff}(p,q)=c(p,q)$ but for non-unitary $h_\text{min}$ is negative.

Characters of the Verma module $\mathcal{V}_{m,n}$:
\begin{equation}\label{eq:char}
    \chi^{(p,q)}_{m,n}(x)=x^{-\frac{c}{24}}\prod_{l=1}^\infty\frac{1}{1-x^l}\sum_{k\in\mathbb{Z}}(x^{h_{m+2kp,n}}-x^{h_{-m+2kp,n}})=x^{h_{m,n}-\frac{c}{24}}\sum\limits_{N=0}^\infty\nu^{(p,q)}_{m,n}(N)x^N\,,
\end{equation}
where $\nu^{(p,q)}_{m,n}(N)$ is the number of linearly independent descendants at level $N$ of the Verma module $\mathcal{V}_{m,n}$.
\newpage
Let us consider the simplest non-unitary minimal model $M(2,5)$ with central charge $c(2,5)=-\frac{22}{5}$ and effective central charge $c_{\text{eff}}(2,5)=\frac{2}{5}$. The GL description of the Yang-Lee model is provided by the scalar field theory with pure imaginary cubic interaction \cite{Fisher:1978pf,Cardy:1985yy}:
\begin{equation}\label{eq12}
    S_{2,5}=\int d^dx\left(\frac{1}{2}(\partial_\mu\phi)^2+\frac{g}{6}\phi^3\right)\,,
\end{equation}
where $g$ is pure imaginary. The action is invariant under $\mathcal{P}\mathcal{T}$ symmetry, which acts by $\phi\rightarrow-\phi$, $i\rightarrow-i$ (also you can see \cite{Bender:2018pbv}). There are $2$ scalar primary operators in $M(2,5)$ $\phi_{1,1}=I$ and $\phi_{1,2}=\phi$, whose properties are listed in Table \ref{tab:M25}.

\begin{table}[htbp]
        \centering
        \begin{tabular}{|c|c|c|c|c|c|c|}
            \hline
            $(2,5)$ & $\phi_{1,1}$ &  $\phi_{1,2}$ \\\hline
            $h_{m,n}$ & $0$ & $-\frac{1}{5}$ \\\hline
            $\mathcal{PT}$ & even & odd \\\hline
        \end{tabular}
        \caption{Primary fields and their properties of $M(2,5)$.}
        \label{tab:M25}
    \end{table}

Characters of Verma modules $\mathcal{V}_{1,1}$ and $\mathcal{V}_{1,2}$:
\begin{equation}\label{eq4}
\begin{aligned}
    \chi^{(2,5)}_{1,1}(x)&=x^{\frac{11}{60}}\prod_{k=1}^\infty\frac{1}{(1-x^{5k-2})(1-x^{5k-3})}=
    \\&=x^{\frac{11}{60}}(1+x^2+x^3+x^4+x^5+2x^6+2x^7+3x^8+3x^9+\mathcal{O}(x^{10}))\,,\\
    \chi^{(2,5)}_{1,2}(x)&=x^{-\frac{1}{60}}\prod_{k=1}^\infty\frac{1}{(1-x^{5k-1})(1-x^{5k-4})}=\\
    &=x^{-\frac{1}{60}}(1+x+x^2+x^3+2x^4+2x^5+3x^6+3x^7+4x^8+5x^9+\mathcal{O}(x^{10}))\,,
\end{aligned}
\end{equation}
where the products can be obtained from (\ref{eq:char}) using Jacobi triple product identity.

Operator Product Expansion (OPE) structure in the Yang-Lee model is simple:
\begin{equation}
\begin{aligned}
    &\phi\times\phi\sim I+i\phi\,,\\
    &\phi(x)\phi(0)=C^I_{\phi\phi}|x|^{\frac{4}{5}}(I+\text{desc.})+C^\phi_{\phi\phi}|x|^{\frac{2}{5}}(\phi(x)+\text{desc.})
\end{aligned}
\end{equation}
with $C^I_{\phi\phi}=1$ and only one non-trivial pure imaginary structure constant $C^\phi_{\phi\phi}=i\kappa$ \cite{Dotsenko:1984nm,Dotsenko:1984ad,Cardy:1985yy}, where
\begin{equation}
    \kappa=\frac{1}{5}\gamma^{\frac{3}{2}}\left(\frac{1}{5}\right)\gamma^{\frac{1}{2}}\left(\frac{2}{5}\right)\approx1.91131\,.
\end{equation}
Here $\gamma(x)=\frac{\Gamma(x)}{\Gamma(1-x)}$.

Consider non-unitary minimal model $M(3,10)$ with central charge $c(3,10)=2c(2,5)=-\frac{44}{5}$ and effective central charge $c_{\text{eff}}(3,10)=2c_{\text{eff}}(2,5)=\frac{4}{5}$. A $D_6$ series version of $M(3,10)=M(2,5)\otimes M(2,5)$, its GL description is provided by two copies of the cubic field theory (\ref{eq12}) with the equal pure imaginary coupling $g$:
\begin{equation}
    S_{3,10}=\int d^dx\left(\frac{1}{2}(\partial_\mu\phi_1)^2+\frac{1}{2}(\partial_\mu\phi_2)^2+\frac{g}{6}(\phi_1^3+\phi_2^3)\right)\,.
\end{equation}
Let us denote $\phi_1=\frac{\sigma+\phi}{\sqrt{2}}$, $\phi_2=\frac{\sigma-\phi}{\sqrt{2}}$ and rewrite the action:
\begin{equation}
    S_{3,10}=\int d^dx\left(\frac{1}{2}(\partial_\mu\phi)^2+\frac{1}{2}(\partial_\mu\sigma)^2+\frac{g_1}{2}\sigma\phi^2+\frac{g_1}{6}\sigma^3\right)\,,
\end{equation}
where $g_1=\frac{g}{\sqrt{2}}$. Here $\mathcal{P}\mathcal{T}$ symmetry acts as $\sigma\rightarrow-\sigma$, $i\rightarrow-i$. Additionally, there is a $\mathbb{Z}_2$ symmetry $\phi\rightarrow-\phi$ that exists for any minimal model $M(p,q)$ if $p$ and $q$ are both not equal to $2$. It is realized by interchanging the two scalar fields $\phi_1$ and $\phi_2$.

A $D_6$ modular invariant partition function:
\begin{equation}\label{PartFunc310}
    Z_{3,10}^{D_6}=|\chi^{(3,10)}_{1,1}+\chi^{(3,10)}_{1,9}|^2+|\chi^{(3,10)}_{1,3}+\chi^{(3,10)}_{1,7}|^2+2|\chi^{(3,10)}_{1,5}|^2\,,
\end{equation}
where characters
\begin{equation}
\begin{aligned}
    &\chi^{(3,10)}_{1,1}(x)=x^{\frac{11}{30}}(1+x^2+x^3+2x^4+2x^5+4x^6+4x^7+7x^8+8x^9+\mathcal{O}(x^{10}))\,,\\
    &\chi^{(3,10)}_{1,3}(x)=x^{-\frac{1}{30}}(1+x+2x^2+2x^3+4x^4+5x^5+8x^6+10x^7+15x^8+19x^9+\mathcal{O}(x^{10}))\,,\\
    &\chi^{(3,10)}_{1,5}(x)=x^{\frac{1}{6}}(1+x+2x^2+3x^3+5x^4+6x^5+10x^6+13x^7+19x^8+25x^9+\mathcal{O}(x^{10}))\,,\\
    &\chi^{(3,10)}_{1,7}(x)=x^{\frac{29}{30}}(1+x+2x^2+3x^3+5x^4+7x^5+10x^6+13x^7+19x^8+25x^9+\mathcal{O}(x^{10}))\,,\\
    &\chi^{(3,10)}_{1,9}(x)=x^{\frac{71}{30}}(1+x+x^2+2x^3+3x^4+4x^5+6x^6+8x^7+11x^8+14x^9+\mathcal{O}(x^{10}))\,.
\end{aligned}
\end{equation}

The partition function of a tensor product $M(2,5)\otimes M(2,5)$:
\begin{align}
    Z^2_{2,5}=(|\chi^{(2,5)}_{1,1}|^2+|\chi^{(2,5)}_{1,2}|^2)^2=|\chi^{(2,5)}_{1,1}|^4+|\chi^{(2,5)}_{1,2}|^4+2|\chi^{(2,5)}_{1,1}|^2|\chi^{(2,5)}_{1,2}|^2\,.
\end{align}
The equity $Z^{D_6}_{3,10}=Z^2_{2,5}$ is equivalent to
\begin{equation}
    \begin{aligned}
        &\chi^{(3,10)}_{1,1}+\chi^{(3,10)}_{1,9}=(\chi^{(2,5)}_{1,1})^2=x^{\frac{11}{30}}\prod_{k=1}^\infty\frac{1}{(1-x^{5k-2})^2(1-x^{5k-3})^2}\,,\\
        &\chi^{(3,10)}_{1,3}+\chi^{(3,10)}_{1,7}=(\chi^{(2,5)}_{1,2})^2=x^{-\frac{1}{30}}\prod_{k=1}^\infty\frac{1}{(1-x^{5k-1})^2(1-x^{5k-4})^2}\,,\\
        &\chi^{(3,10)}_{1,5}=\chi^{(2,5)}_{1,1}\chi^{(2,5)}_{1,2}=x^{\frac{1}{6}}\prod_{k=1}^\infty\frac{1}{(1-x^{5k-1})(1-x^{5k-2})(1-x^{5k-3})(1-x^{5k-4})}\,.
    \end{aligned}
\end{equation}

Partition function (\ref{PartFunc310}) can be obtained by orbifolding the $\mathbb{Z}_2$ symmetry in the $A$ modular invariant. The $\mathbb{Z}_2$ orbifold keeps the even sector, removes the odd one but also adds the twisted sector \cite{DiFrancesco:1997nk}. All fields from the even sector are $\mathbb{Z}_2$ even, and from the twisted are $\mathbb{Z}_2$ odd in a $D$ series.

The primary field content of the $D_6$ modular invariant $M(3,10)$ \cite{Kausch:1996vq}:
\begin{itemize}
    \item Even sector (5 fields): $\phi_{1,1}$, $\phi_{1,3}$, $\phi_{1,5}^+$, $\phi_{1,7}$, $\phi_{1,9}$;
    \item Twisted sector (5 fields): $\phi_{1,1}\bar{\phi}_{1,9}$, $\phi_{1,3}\bar{\phi}_{1,7}$, $\phi_{1,5}^-$, $\phi_{1,7}\bar{\phi}_{1,3}$, $\phi_{1,9}\bar{\phi}_{1,1}$.
\end{itemize}

There are $6$ scalar primary operators in a $D_6$ series (whole even sector and one field from the twisted one), whose properties are listed in Table 1 of \cite{Katsevich:2024jgq}. They can be written in terms of primary fields of $M(2,5)$:
\begin{equation}
    \begin{aligned}
        \phi_{1,1}&=I\otimes I\,,\\
        \phi_{1,3}&=\phi\otimes\phi\,,\\
        \phi^+_{1,5}&=I\otimes\phi+\phi\otimes I\,,\\
        \phi^-_{1,5}&=I\otimes\phi-\phi\otimes I\,,\\
        \phi_{1,7}&=L_{-1}\bar{L}_{-1}\phi\otimes\phi+\phi\otimes L_{-1}\bar{L}_{-1}\phi-L_{-1}\phi\otimes\bar{L}_{-1}\phi-\bar{L}_{-1}\phi\otimes L_{-1}\phi\,,\\
        \phi_{1,9}&=L_{-2}\bar{L}_{-2}I\otimes I+I\otimes L_{-2}\bar{L}_{-2}I-L_{-2}I\otimes\bar{L}_{-2}I-\bar{L}_{-2}I\otimes L_{-2}I\,.
    \end{aligned}
\end{equation}

Recently, it was proposed that the GL description of the whole $D$ series $M(q,3q\pm1)$ for odd $q\geq 3$ is given by a two-field action \cite{Katsevich:2024jgq}:
\begin{align}
    S_{q,3q\pm1}=\int d^dx\left(\frac{1}{2}(\partial_\mu\phi)^2+\frac{1}{2}(\partial_\mu\sigma)^2+\sum_{j=1}^{\frac{q+1}{2}}\frac{g_j}{(2j-1)!(q-2j+1)!}\sigma^{2j-1}\phi^{q-2j+1}\right)\,,
\end{align}
where all $g_i$ are pure imaginary. 

\section{Theoretical predictions for spectrum}
\label{sec:predictions}
\subsection{\texorpdfstring{$M(2,2n+1)$}{M(2,2n+1)} perturbed by \texorpdfstring{$\phi_{1,3}$}{}}
Some reductions of the sine-Gordon models describe $\phi_{1,3}$-perturbations of minimal models. The correspondence between the sine-Gordon model and perturbed minimal models $M(2,2n+1)+\phi_{1,3}$ was investigated in \cite{Smirnov:1990vm}. 

Consider the sine-Gordon model with action:
\begin{equation}
    S_{SG}=\int d^2x\left(\frac{(\partial_\mu\phi)^2}{8\pi}+M\cos\beta\phi\right)\,.
\end{equation}

Let us divide action as
\begin{equation}
    S_{SG}=S_0+S_1\,,
\end{equation}
where
\begin{equation}
    S_0=\int d^2x\left(\frac{(\partial_\mu\phi)^2}{8\pi}+\frac{M}{2}e^{-i\beta\phi}\right),\quad S_1=\frac{M}{2}\int d^2xe^{i\beta\phi}\,.
\end{equation}

The first term $S_0$ describes a model with central charge
\begin{equation}\label{eq9}
    c_{SG}=1-6\left(\frac{\sqrt{2}}{\beta}-\frac{\beta}{\sqrt{2}}\right)^2\,.
\end{equation}
On the other hand, this is a central charge of $M(2,2n+1)$:
\begin{equation}\label{eq10}
    c(2,2n+1)=1-\frac{3(1-2n)^2}{1+2n}\,.
\end{equation}
Comparing (\ref{eq9}) and (\ref{eq10}), we can obtain
\begin{equation}
    \beta^2=\frac{4}{1+2n}<1\,.
\end{equation}

The second term $S_1$ is perturbation. Its holomorphic dimension:
\begin{equation}
    h=\beta\left(\beta-\frac{1}{\beta}\right)=\frac{1}{4}\left(\frac{3\sqrt{2}}{\beta}-\frac{\beta}{\sqrt{2}}\right)^2-\frac{1}{4}\left(\frac{\sqrt{2}}{\beta}-\frac{\beta}{\sqrt{2}}\right)^2=h_{13}\,.
\end{equation}

Spectrum of sine-Gordon theory consists of two types of particles: solitons and breathers. Breathers are boundary states of kink (one-soliton solution) with anti-kink. Breather masses:

\begin{equation}
    m_k=2m_\text{kink}\sin\frac{\pi pk}{2}\,,
\end{equation}
where $k\in\mathbb{N}\cap\left[0,\frac{1}{p}\right)$ and $p$ can be found from the equation
\begin{equation}
    \beta^2=\frac{2p}{p+1}\,.
\end{equation}
For perturbed models $M(2,2n+1)$:
\begin{align}
    p=\frac{2}{2n-1}\,.
\end{align}
There are no kinks in a series of perturbed $M(2,2n+1)+\phi_{1,3}$, only breathers with masses \cite{Freund:1989jq}:
\begin{align}
    m_k=m_1\frac{\sin\frac{\pi k}{2n-1}}{\sin\frac{\pi}{2n-1}}\,,
\end{align}
where $k\in\{1,2,...,n-1\}$.

There is a similar result for minimal model $M(2,2n+1)$ perturbed by $\phi_{1,2}$ \cite{Koubek:1991dt,Smirnov:1991uw}:
\begin{align}
    m_k=m_1\frac{\sin\frac{\pi k}{6n}}{\sin\frac{\pi}{6n}}\,.
\end{align}

Let us apply this to the SLYM with $p=\frac{2}{3}$. This theory contains only one particle (breather) with mass $m_1$ and no kinks. Scaling Lee-Yang Model corresponds to a pure imaginary coupling $\lambda=ih$ of the field $\phi$ with the action \cite{Cardy:1985yy}:
\begin{equation}
    S=S_{2,5}+ih\int d^2x\phi(x)\,.
\end{equation}
The GL description of SLYM:
\begin{equation}
    S=\int d^dx\left(\frac{1}{2}(\partial_\mu\phi)^2+ih\phi+\frac{g}{6}\phi^3\right)\,.
\end{equation}
In two-dimensional case, from the dimensional analysis $\lambda=\lambda_0m^{12/5}$, where $\lambda_0$ is dimensionless. SLYM is integrable, its spectrum contains a single neutral particle (breather) of mass $m$ \cite{Cardy:1989fw}. The relation between the mass and coupling constant \cite{Zamolodchikov:1989cf,Zamolodchikov:1995xk}:
\begin{equation}
    m=Ch^{5/12},\quad C=\frac{2^{\frac{19}{12}}\sqrt{\pi}(\Gamma(\frac{3}{5})(\Gamma(\frac{4}{5}))^{\frac{5}{12}}}{5^{\frac{5}{16}}\Gamma(\frac{2}{3})\Gamma(\frac{5}{6})}\approx2.64294\,.
\end{equation}

The vacuum energy density \cite{Zamolodchikov:1989cf,Destri:1990ps}:
\begin{equation}
    F=fm^2,\quad f=-\frac{\sqrt{3}}{12}\,.  
\end{equation}
\subsection{\texorpdfstring{$M(3,10)$}{M(3,10)} perturbed by \texorpdfstring{$\phi_{1,3}$}{}}\label{phi13}
There is no evidence about the spectrum of the minimal model $M(3,10)$ perturbed by field $\phi_{1,3}$. $M(3,10)$ has $\mathbb{Z}_2$ symmetry. Perturbing it by $\mathbb{Z}_2$-even operator $\phi_{1,3}$, we can make this $\mathbb{Z}_2$ symmetry spontaneously broken. Then there must be a kink solution.

Consider this perturbed CFT semiclassically\footnote{The semiclassical description was developed in discussions with Igor Klebanov.}. The action of minimal model $M(3,10)$ perturbed by $\phi_{1,3}$ with real coupling $\lambda$:
\begin{equation}
    S=S_{3,10}+\lambda\int d^2x\phi_{1,3}(x)\,.
\end{equation}
The field $\phi_{1,3}$ is the $\mathbb{Z}_2$-even operator $\phi_1\phi_2$. The GL description:
\begin{equation}
    S=\int d^dx\left(\frac{1}{2}(\partial_\mu\phi_1)^2+\frac{1}{2}(\partial_\mu\phi_2)^2+\frac{g}{6}(\phi_1^3+\phi_2^3)+\lambda\phi_1\phi_2\right)\,.
\end{equation}
Let us denote $\phi_1=\frac{\sigma+\phi}{\sqrt{2}}$, $\phi_2=\frac{\sigma-\phi}{\sqrt{2}}$ and rewrite the action:
\begin{equation}
    S=\int d^dx\left(\frac{1}{2}(\partial_\mu\phi)^2+\frac{1}{2}(\partial_\mu\sigma)^2+\frac{g_1}{2}\sigma\phi^2+\frac{g_1}{6}\sigma^3+\frac{\lambda}{2}\sigma^2-\frac{\lambda}{2}\phi^2\right)\,,
\end{equation}
where $g_1=\frac{g}{\sqrt{2}}\in i\mathbb{R}$. The semiclassical potential:
\begin{equation}
    V(\sigma,\phi)=\frac{\lambda}{2}\phi^2-\frac{\lambda}{2}\sigma^2-\frac{g_1}{2}\sigma\phi^2-\frac{g_1}{6}\sigma^3\,,
\end{equation}
where fields $\sigma\in i\mathbb{R},\phi\in\mathbb{R}$ because the physical potential should be real. Replace $\sigma\rightarrow i\sigma$:
\begin{equation}
    V(i\sigma,\phi)=\frac{\lambda}{2}\phi^2+\frac{\lambda}{2}\sigma^2-\frac{ig_1}{2}\sigma\phi^2+\frac{ig_1}{6}\sigma^3\,.
\end{equation}
Equations of motion:
\begin{equation}\label{eqofmotion}
\begin{cases}
    \Box\phi=-\lambda\phi+ig_1\sigma\phi\,,\\
    \Box\sigma=\lambda\sigma-\frac{ig_1}{2}\phi^2+\frac{ig_1}{2}\sigma^2\,.
\end{cases}
\end{equation}
Stationary points:
\begin{equation}
    (\sigma_0,\phi_0)=\frac{i\lambda}{g_1}\begin{cases}
        (0,0)\,,\\
        (-1,\pm\sqrt{3})\,,\\
        (2,0)\,.
    \end{cases}
\end{equation}
Boundary conditions:
\begin{equation}
    \sigma(-\infty)=\sigma(\infty)=-\frac{i\lambda}{g_1},\quad\phi(-\infty)=-\frac{\sqrt{3}i\lambda}{g_1},\quad\phi(+\infty)=\frac{\sqrt{3}i\lambda}{g_1}\,.
\end{equation}
\begin{equation}
    V\left(\frac{\lambda}{g_1},\pm\frac{\sqrt{3}i\lambda}{g_1}\right)=\frac{2\lambda^3}{3|g_1|^2}\,.
\end{equation}

System with another rhs in the second equation 
\begin{equation}
    \begin{cases}
    \Box\phi=-\lambda\phi+ig_1\sigma\phi\,,\\
    \Box\sigma=\lambda\sigma-\frac{ig_1}{2}\phi^2+6\frac{ig_1}{2}\sigma^2\,.
    \end{cases}
\end{equation}
has analytic stationary kink one-dimensional solution
\begin{equation}
    \phi(x)=\pm\frac{2\sqrt{2}i\lambda\tanh(C_1x)}{g_1},\quad\sigma(x)=\mp\frac{C_1}{\sqrt{2}\lambda}\phi'(x)-\frac{i\lambda}{g_1}\,,
\end{equation}
where $C_1$ can be obtained from the biquadratic equation:
\begin{equation}
    4C_1^4+7C_1^2\lambda+2\lambda^2=0\leftrightarrow C_1=\pm\frac{\sqrt{(-7\pm\sqrt{17})\lambda}}{2\sqrt{2}}\,.
\end{equation}
Maybe, our system (\ref{eqofmotion}) also has a similar solution. It would be interesting to find it numerically.

Using intuition that $M(3,10)$ is strongly correlated with the minimal model $M(2,5)$ (as tensor square), we can assume that the theory from the previous subsection can be relevant for $M(3,10)$. Let us substitute the central charge $c(3,10)=-\frac{44}{5}$ in eq. (\ref{eq9}):
\begin{equation}
    \beta^2=\frac{2p}{p+1}=\frac{3}{5}\leftrightarrow p=\frac{3}{7}\,.
\end{equation}
Thus, our conjecture is that \textit{perturbed minimal model $M(3,10)+\phi_{1,3}$ contains kink $m_\text{kink}$ and two breathers with masses}:
\begin{equation}\label{conj}
    \begin{aligned}
        m_1=2m_\text{kink}\sin\frac{3\pi}{14}\approx1.247m_\text{kink}\,,\\
        m_2=2m_\text{kink}\sin\frac{3\pi}{7}\approx1.950m_\text{kink}\,.
    \end{aligned}
\end{equation}
It will be supported by TCSA in section \ref{TCSA1}.
\subsection{\texorpdfstring{$M(3,10)$}{M(3,10)} perturbed by \texorpdfstring{$i\phi^+_{1,5}$}{}}\label{phi15}
Minimal model $M(3,10)$ perturbed by $\phi^+_{1,5}$ was described in \cite{Kausch:1996vq}. The field $\phi^+_{1,5}$ is the $\mathbb{Z}_2$-even operator $\phi_1+\phi_2$. There is an identity
\begin{equation}
    M(3,10)+i\phi^+_{1,5}=(M(2,5)+i\phi)\otimes(M(2,5)+i\phi)\,.
\end{equation}
The spectrum consists of two particles (breathers) of equal mass $m$. It will be supported by TCSA in section \ref{TCSA2}.

Also, this RG flow can be considered as a massive $k=2,I=4$ flow from \cite{Nakayama:2024msv}:
\begin{equation}
    M(3,10)+i\phi^+_{1,5}\rightarrow M(2,3)\,.
\end{equation}

\section{Truncated Conformal Space Approach}
\label{sec:TCSA}
\subsection{Perturbed CFT on a cylinder}
Consider two-dimensional field theory on a cylinder (see, for example, \cite{Yurov:1989yu,Zamolodchikov:1989hfa}). Let us denote the compact coordinate as $x\in[0,R]$, $R$ is circumference, and non-compact as $y\in(-\infty,\infty)$. The coordinate $x$ plays a role of spatial coordinate and $y$ plays a role of time. The complex coordinates:
\begin{equation}
    \zeta=x+iy,\quad\Bar{\zeta}=x-iy\,.
\end{equation}
Exponential map on a plane:
\begin{equation}
    z=e^{2\pi i\frac{\zeta}{R}},\quad \bar{z}=e^{-2\pi i\frac{\bar{\zeta}}{R}}\,.
\end{equation}
The scaling dimension of $\phi$ is $h+\bar{h}=2h=\Delta$. Action is dimensionless ($\hbar=1$), so the coupling $\lambda$ can be expressed as
\begin{equation}
    \lambda=\lambda_0m^{2-\Delta}\,,
\end{equation}
where $\lambda_0$ is dimensionless. Primary field $\phi$ is relevant when its holomorphic dimension $h<1$. To avoid UV divergences, we have to put $h<\frac{1}{2}$. In the general case, perturbation may consist of an arbitrary number of relevant fields.

The Hamiltonian of perturbed CFT:
\begin{equation}\label{eq2}
    H=H_{CFT}+V\,.
\end{equation}
Unperturbed part on a cylinder of radius $R$:
\begin{equation}
    H_{CFT}=\frac{2\pi}{R}\left(L_0+\Bar{L}_0-\frac{c}{12}\right)\,.
\end{equation}
Perturbation on a cylinder:
\begin{equation}\label{eq3}
    V=\lambda\int_0^R dx\phi(x,0)\,.
\end{equation}

Let us consider the momentum operator:
\begin{equation}
    P=\frac{2\pi}{R}(L_0-\bar{L}_0)\,.
\end{equation}
Hamiltonian (\ref{eq2}) commutes with it $[H,P]=0$, so the Hilbert space factorizes on sectors with fixed momentum
\begin{equation}
    P=\frac{2\pi}{R}s\,,
\end{equation}
where $s=L_0-\bar{L}_0\in\mathbb{Z}$ is a spin.

Integration by spatial coordinate in (\ref{eq3}) gives the conservation of spin in matrix elements of perturbation $V$:
\begin{equation}
    \braket{h_\beta|V|h_\alpha}=\lambda R\delta_{s_\alpha,s_\beta}\braket{h_\beta|\phi(0,0)|h_\alpha}\,,
\end{equation}
where $\ket{h_\alpha}$ is an arbitrary descendant of primary field $\ket{\phi_\alpha}$: $\ket{h_\alpha}=L_{-\bm{\mu}}\ket{\phi_\alpha}\equiv L_{-\mu_1}L_{-\mu_2}...\ket{\phi_\alpha}$ ($\mu_1\geq\mu_2\geq...$).
Let us introduce the matrix element:
\begin{equation}
    \braket{h_\beta|V|h_\alpha}=\braket{h_\beta|L_{\bm{\mu}}\phi(z)L_{-\bm{\lambda}}|h_\alpha}\equiv\lim\limits_{\zeta\rightarrow\infty}|\zeta|^{\Delta_\beta}\braket{\phi^{\bm{\lambda}}_\alpha(1,1)\phi(z,\bar{z})\phi^{\bm{\mu}}_\beta(\zeta,\bar{\zeta})}\,,
\end{equation}
where $\phi^{\bm{\lambda}}_{\alpha}=L_{-\bm{\lambda}}\phi_{\alpha}$. The three-point function between primary fields $\phi_1$, $\phi_2$ and $\phi_3$:
\begin{equation}
    \braket{\phi_1(z_1)\phi_2(z_2)\phi_3(z_3)}=C_{\phi_1\phi_2\phi_3}\prod_{i<j}(z_i-z_j)^{-h_{ij}}\,,
\end{equation}
where $h_{12}=h_1+h_2-h_3$, $h_{13}=h_1+h_3-h_2$, $h_{23}=h_2+h_3-h_1$ and $C_{\phi_1\phi_2\phi_3}$ are structure constants. Matrix elements between two primary fields are proportional to the structure constants:
\begin{equation}
    \braket{\phi_\beta|\phi(0)|\phi_\alpha}=\left(\frac{2\pi}{R}\right)^{\Delta}C_{\phi_\beta\phi\phi_\alpha}\,.
\end{equation}
An arbitrary matrix element is computed with the help of commutation relations of Virasoro algebra and
\begin{equation}
\begin{aligned}
    \relax [L_n,\phi(z,\bar{z})]=z^n(z\partial_z+h(n+1))\phi(z,\bar{z})\,,\\
    [\bar{L}_n,\phi(z,\bar{z})]=\bar{z}^n(\bar{z}\partial_{\bar{z}}+
    \bar{h}(n+1))\phi(z,\bar{z})\,.
\end{aligned}
\end{equation}
Denote differential operator $\mathcal{L}_n=z^n(z\partial_z+h(n+1))$. We have $[L_n,\phi(z,\bar{z})]=\mathcal{L}_n\cdot\phi(z,\bar{z})$ and
\begin{equation}
    [L_{\lambda_n},[L_{\lambda_{n-1}},...,[L_{\lambda_1},\phi(z,\bar{z})]]]=\mathcal{L}_{\lambda_1}\cdot...\cdot\mathcal{L}_{\lambda_n}\phi(z,\Bar{z})\equiv\mathcal{L}_{\bm{\lambda}}\phi(z,\bar{z})\,.
\end{equation}
Thus, we obtain that for arbitrary basis states $\ket{h_{\alpha,\beta}}$ of the Hilbert space:
\begin{equation}
    \braket{h_\beta|\phi(0)|h_\alpha}=\left(\frac{2\pi}{R}\right)^{\Delta}B_{\alpha\beta}\,,
\end{equation}
where $B_{\alpha\beta}\propto C_{\phi_\beta\phi\phi_\alpha}$ are dimensionless coefficients. Later we will demonstrate a useful basis of quasi-primary fields and give a receipt of calculation of $B_{\alpha\beta}$. Matrix element of a full Hamiltonian:
\begin{equation}
    H_{\alpha\beta}=\frac{2\pi}{R}\left(\left(\Delta_\alpha-\frac{c}{12}\right)\delta_{\alpha\beta}+G\delta_{S_\alpha,S_\beta}B_{\alpha\beta}\right)\,,
\end{equation}
where $G=\lambda(2\pi)^{\Delta-1}R^{2-\Delta}=\lambda_0(2\pi)^{\Delta-1}r^{2-\Delta}$ is a dimensionless effective coupling constant, $r=mR$ is the scaling length.

Perturbed Hamiltonian in a concrete Verma module ($S_\alpha=S_\beta$):
\begin{equation}\label{eq8}
    H=\frac{2\pi}{R}\left(\Delta-\frac{c}{12}\right)\bm{1}+\frac{2\pi G}{R}B\,.
\end{equation}
\subsection{Level and energy truncation}
Usually, people consider two ways of Truncated Approach realization:
\begin{itemize}
    \item Level truncation. That means truncation of Hilbert space up to level $n$ in every Verma module. The maximal value of energy operator $L_0+\bar{L}_0$ is $h+\Bar{h}+2n$. Overheight representations with high $\Delta$ can have a big impact on spectrum. 
    \item Energy truncation. That means truncation in the way to do $L_0 + \bar{L}_0 \leq 2n$. In this method representations with high $\Delta$ have not an excessive influence on spectrum but, for example, representations with $\Delta > n$ do not count at all.
\end{itemize}
These methods give similar results when $n\gg\Delta$ for any $\Delta$ in the model. We will apply the level truncation for $M(2,5)$ and the energy truncation for $M(3,10)$.

The dimension of spin $s$ sector of the minimal model $M(p,q)$ which is truncated up to level $N$:
\begin{equation}\label{eq11}
    \mathcal{N}^{(p,q)}(N,s)=\sum\limits_{\Delta}\sum\limits_{i=0}^{N-s}\nu^{(p,q)}_\Delta(i)\nu^{(p,q)}_\Delta(i+s)\,,
\end{equation}
where $\Delta$ lists conformal families in the model.

For example, for the Yang-Lee model $M(2,5)$:
\begin{equation}
\begin{aligned}
    \mathcal{N}^{(2,5)}(5,0)&=5+12=17\,,\\
    \mathcal{N}^{(2,5)}(15,0)&=280+677=957\,.
\end{aligned}
\end{equation}

To calculate the matrix $B$ in (\ref{eq8}) we should choose the basis of states in the Hilbert space.

\subsection{Basis of quasi-primary fields and their derivatives}
A quasi-primary field $\ket{h}$ is a field that satisfies the condition $L_1\ket{h}=0$. It cannot be rewritten as a derivative of the field from the previous level.

Space $\mathcal{V}_{h,N}=\text{span}\{L_{-\bm{\lambda}}\ket{h}:|\bm{\lambda}|\equiv\lambda_1+\lambda_2+...=N\}$ can be decomposed into two orthogonal subspaces: the first subspace consists of quasi-primary fields at level $N$, the second consists of derivatives $L_{-1}^m\ket{h_{N-m}}$, where $\ket{h_{N-m}}$ is a quasi-primary field at level $N-m$. So the number of linearly independent quasi-primary fields at level $N$ is
\begin{equation}\label{eq6}
    Q_\Delta(N)=\nu_\Delta(N)-\nu_\Delta(N-1)\,.
\end{equation}

From the character $\chi_{1,1}(x)$ (\ref{eq4}) the dimension $\nu_I(N)$ and number of quasi-primary states (\ref{eq6})
\begin{equation}
    Q_I(N)=\nu_I(N)-\nu_I(N-1)+\delta_{N,1}
\end{equation}
at level $N$ for the Verma module $I$ are listed in table \ref{tab1}. We put $\nu_I(-1)=0$, the term $\delta_{N,1}$ is added assuming that $L_{-1}I=0$.
\begin{table}[htbp]
    \centering
    \begin{tabular}{|c|c|c|c|c|c|c|c|c|c|c|c|c|c|c|c|c|}
    \hline
        $N$ & 0 & 1 & 2 & 3 & 4 & 5 & 6 & 7 & 8 & 9 & 10 & 11 & 12 & 13 & 14 & 15\\\hline
        $\nu_I(N)$ & 1 & 0 & 1 & 1 & 1 & 1 & 2 & 2 & 3 & 3 & 4 & 4 & 6 & 6 & 8 & 9\\\hline
        $Q_I(N)$ & 1 & 0 & 1 & 0 & 0 & 0 & 1 & 0 & 1 & 0 & 1 & 0 & 2 & 0 & 2 & 1 \\\hline
    \end{tabular}
    \caption{The dimension and number of quasi-primary fields of the Verma module $I$}
    \label{tab1}
\end{table}

From the character $\chi_{1,2}(x)$ (\ref{eq4}) the dimension $\nu_I(N)$ and number of quasi-primary states (\ref{eq6})
\begin{equation}
    Q_\phi(N)=\nu_\phi(N)-\nu_\phi(N-1)
\end{equation}
at level $N$ for the Verma module $\phi$ are listed in table \ref{tab2}. We put $\nu_\phi(-1)=0$.

\begin{table}[htbp]
    \centering
    \begin{tabular}{|c|c|c|c|c|c|c|c|c|c|c|c|c|c|c|c|c|}
    \hline
        $N$ & 0 & 1 & 2 & 3 & 4 & 5 & 6 & 7 & 8 & 9 & 10 & 11 & 12 & 13 & 14 & 15\\\hline
        $\nu_\phi(N)$ & 1 & 1 & 1 & 1 & 2 & 2 & 3 & 3 & 4 & 5 & 6 & 7 & 9 & 10 & 12 & 14\\\hline
        $Q_\phi(N)$ & 1 & 0 & 0 & 0 & 1 & 0 & 1 & 0 & 1 & 1 & 1 & 1 & 2 & 1 & 2 & 2 \\\hline
    \end{tabular}
    \caption{The dimension and number of quasi-primary fields of the Verma module $\phi$}
    \label{tab2}
\end{table}

Basis of quasi-primary fields is useful for computation because the matrix element between two derivates $\braket{h_\beta|L_1^k\phi(0)L_{-1}^n|h_\alpha}$ can be algebraically expressed via matrix element between quasi-primary fields $\braket{h_\beta|\phi(0)|h_\alpha}$ \cite{Yurov:1989yu}:
\begin{equation}
\begin{aligned}
    &\braket{h_\beta|L_1^k\phi(0)L_{-1}^n|h_\alpha}=\\
    &=k!n!\sum\limits_{l=0}^{\min(k,n)}\frac{(h_\beta+h_\alpha-h)_l(h+h_\alpha-h_\beta)_{n-l}(h+h_\beta-h_\alpha)_{k-l}}{l!(k-l)!(n-l)!}\braket{h_\beta|\phi(0)|h_\alpha}\,.
\end{aligned}
\end{equation}
Norm of the derivative $L_{-1}^n\ket{h}$:
\begin{equation}\label{eq7}
    \braket{h|L_1^nL_{-1}^n|h}=n!\prod\limits_{j=0}^{n-1}(2h+j)=n!(2h)_n\,,
\end{equation}
where $(x)_n=\frac{\Gamma(x+n)}{\Gamma(x)}$ is a Pochhammer symbol.

Basis of spin $s$ fields consists of 
\begin{equation}
    L_{-1}^{k_1}\bar{L}_{-1}^{k_2}\ket{h_\alpha},\quad s=k_1-k_2\,,
\end{equation}
where $\ket{h_\alpha}$ are quasi-primary fields. These fields should be normalized using (\ref{eq7}).

List of quasi-primary fields we have calculated up to the level 15 in Appendix \ref{qfields}. Another list of quasi-primary fields up to the level 12 is given in propositions 4,5 of \cite{Leitner:2018iyf}.

In principle, we may not use a basis of quasi-primary fields, just making a basis with all possible descendants and reducing all null-vectors.

New effective algorithms in Wolfram Mathematica for obtaining null-vectors, Gram matrices and quasi-primary fields in Wolfram Mathematica can be found in \cite{Fitzpatrick:2023aqm}.
\subsection{Basis in spin-\texorpdfstring{$0$}{0} sector of \texorpdfstring{$M(3,10)$}{M(3,10)}}
Let us build a basis of spin-$0$ sector. Using the fact that a $D_6$ series version of $M(3,10)=M(2,5)\otimes M(2,5)$, the basis consists of fields
\begin{equation}
    L_{-1}^{k_1}\bar{L}_{-1}^{k_2}\ket{h_\alpha}\otimes L_{-1}^{l_1}\bar{L}_{-1}^{l_2}\ket{h_\beta}\,, 
\end{equation}
where $k_1$, $k_2$, $l_1$, $l_2$ are integers. Spin $s=0$ condition:
\begin{equation}
    s=k_1-k_2+l_1-l_2=0\,.
\end{equation}

Unperturbed Hamiltonian:
\begin{equation}
    H_{CFT}=\left(L_0+\bar{L}_0-\frac{c(2,5)}{12}\right)\otimes1+1\otimes\left(L_0+\bar{L}_0-\frac{c(2,5)}{12}\right)\,.
\end{equation}
Corresponding energy:
\begin{equation}
    \mathcal{E}_{\alpha,\beta,k_1,k_2,l_1,l_2}=\Delta_\alpha+k_1+k_2+\Delta_\beta+l_1+l_2-\frac{c(2,5)}{6}\,.
\end{equation}

We will do energy truncation because the level of the tensor product of states is badly defined. Size of matrix $H$ depending on the truncation energy $\mathcal{E}$ is represented in Table \ref{tab4}.
\begin{table}[htbp]
        \centering
        \begin{tabular}{|c|c|c|c|c|c|}
    \hline
        $\mathcal{E}$ & 5 & 7 & 9 & 11 & 13\\\hline
        $\mathcal{N}^{(3,10)}(\mathcal{E}, 0)$ & 31 & 69 & 177 & 365 & 839 \\\hline
    \end{tabular}
        \caption{Dimension of spin-$0$ sector truncated up to the energy $\mathcal{E}$}
        \label{tab4}
    \end{table}
\section{Application of Truncated Approach}
\label{sec:Application}
TCSA was done in Wolfram Mathematica 14.0. All code and results of its work (matrices and plots) can be found at \href{https://github.com/Andrew-Kot/TCSAM310/tree/main}{GitHub} \footnote{A part of the code is taken from the Matthew Headrick's WM package \href{https://sites.google.com/view/matthew-headrick/mathematica?authuser=0}{Virasoro.nb}.}.
\subsection{Scaling Lee-Yang Model}
In a pioneer work \cite{Yurov:1989yu} TCSA was applied for SLYM up to the level 5 with 17 basis states in the case of spin $s=0$. In this subsection we will apply TCSA up to the level 15 with 957 states in the case of spin $s=0$ reproducing well-known results with more accuracy.
\subsubsection{Spin \texorpdfstring{$s=0$}{s=0} sector}
Consider Hilbert space with states of spin $s=0$. Size of matrix $H$ depending on truncation level $N$ is calculated using (\ref{eq11}) and represented in the Table \ref{tab3}.
\begin{table}[htbp]
        \centering
        \begin{tabular}{|c|c|c|c|c|c|c|}
    \hline
        $N$ & 5 & 7 & 9 & 11 & 13 & 15\\\hline
        $\mathcal{N}^{(2,5)}(N,0)$ & 17 & 43 & 102 & 219 & 472 & 957\\\hline
    \end{tabular}
        \caption{Dimension of spin $s=0$ sector truncated up to the level $N$}
        \label{tab3}
\end{table}

\begin{figure}[htbp]
\minipage{0.32\textwidth}
  \includegraphics[width=\linewidth]{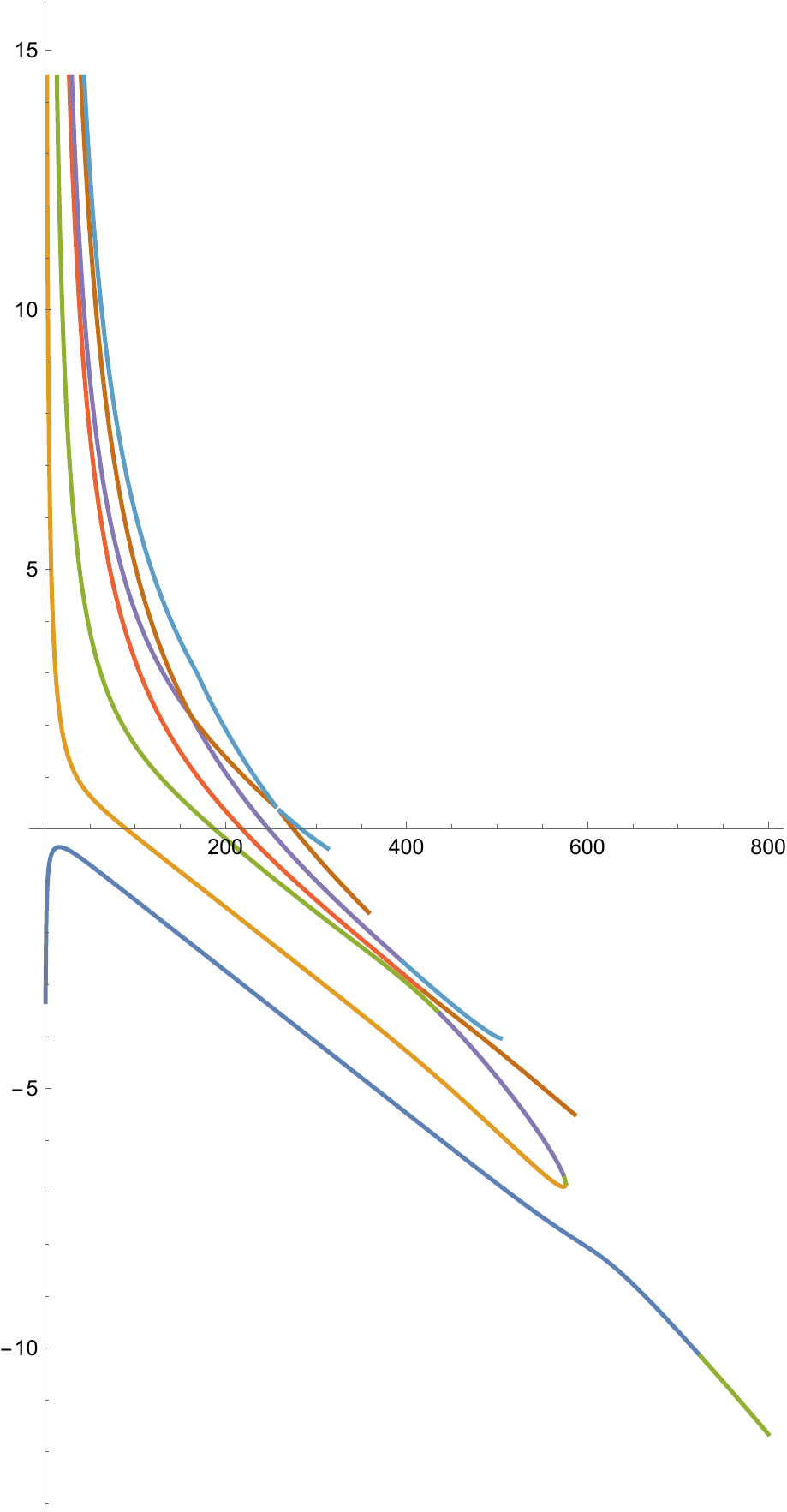}
  \caption{Energy levels $E_i(r)$ in spin $s=0$ sector. Truncation was made up to the level $N=5$.}\label{fig:1}
\endminipage\hfill
\minipage{0.32\textwidth}
  \includegraphics[width=\linewidth]{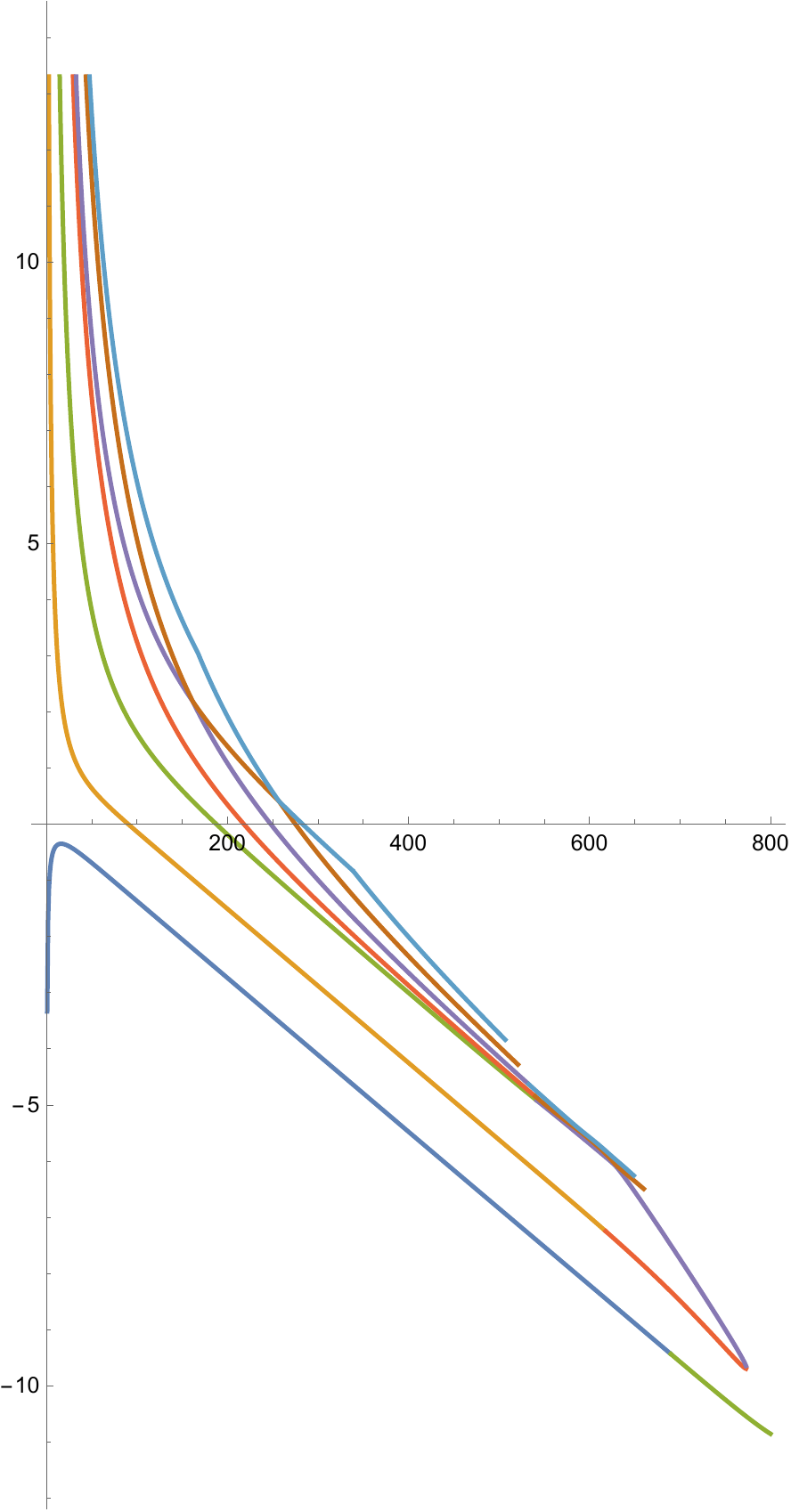}
  \caption{Energy levels $E_i(r)$ in spin $s=0$ sector. Truncation was made up to the level\\ $N=11$.}\label{fig:2}
\endminipage\hfill
\minipage{0.32\textwidth}%
  \includegraphics[width=\linewidth]{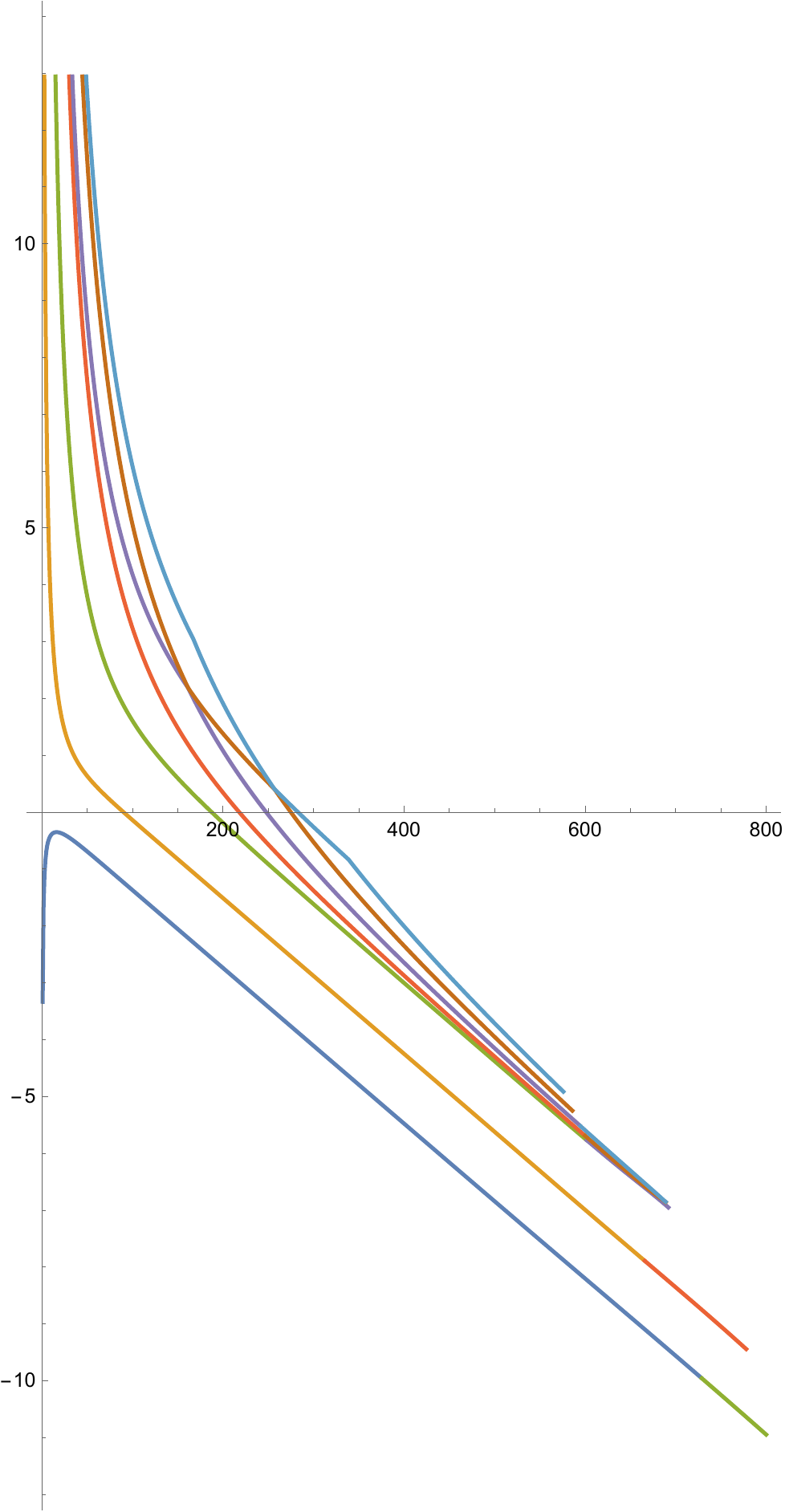}
  \caption{Energy levels $E_i(r)$ in spin $s=0$ sector. Truncation was made up to the level\\ $N=15$.}\label{fig:3}
\endminipage
\end{figure}

Using the TCSA algorithm, we can calculate the matrix elements of Hamiltonian $H$ for spin $s=0$ and different truncation levels $N$. The first 7 energy levels of $H$ against the scaling length $r$ for $N=5,11,15$ are plotted in Fig. \ref{fig:1}, \ref{fig:2}, \ref{fig:3}. The Fig. \ref{fig:1} with $N=5$ is analogous to Fig. 2 in \cite{Yurov:1989yu}.

\begin{figure}[htbp]
    \centering
    \includegraphics[scale=0.5]{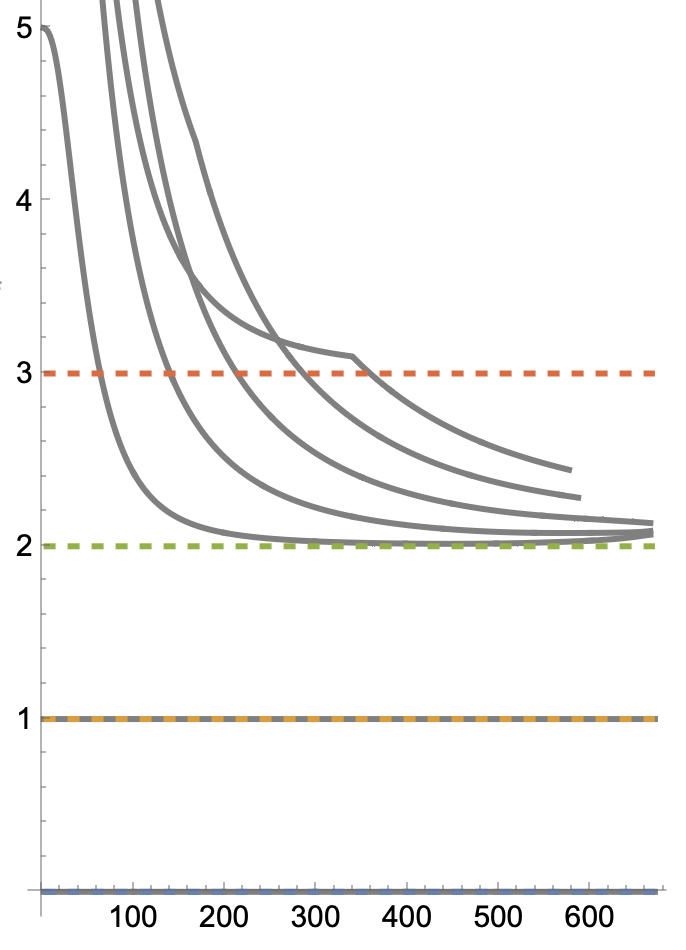}
    \caption{$\frac{E_i-E_0}{E_1-E_0}(r)$ in spin $s=0$ sector. Truncation was made up to the level $N=15$. Dashed blue corresponds to vacuum, yellow to one-breather state $m$, green to two-breather states $2m$, red to three-breather states $3m$.}\label{fig:13}
\end{figure}

The first 7 levels of the spectrum at large $r\rightarrow\infty$: the ground state $\phi$ (without particles), one-particle state $I$, three two-particle states $\partial\bar{\partial}\phi$, $\partial^2\Bar{\partial}^2\phi$ and $\partial^3\Bar{\partial}^3\phi$, two three-particle states $T\Bar{T}$ and $\partial T\bar{\partial}\Bar{T}$. The 5th and 6th levels corresponding to $\partial^3\Bar{\partial}^3\phi$ and $T\Bar{T}$ are intersecting. The plot of $\frac{E_i-E_0}{E_1-E_0}$, where $i\in\{1,...,7\}$ is the corresponding energy level, against the scaling length $r$ for $N=15$ is depicted in Fig. \ref{fig:13}.

Matrix elements of perturbation $B$ can be useful to refine the results of the numerical solution for the reduced partition function on a sphere in \cite{Zamolodchikov:2001dz}.


\newpage
\subsubsection{Spin \texorpdfstring{$s=1$}{s=1} sector}
Consider Hilbert space with states of spin $s=1$. The size of matrix $H$ depending on truncation level $N$ is calculated using (\ref{eq11}) and represented in the Table \ref{tab5}.
\begin{table}[htbp]
        \centering
        \begin{tabular}{|c|c|c|c|c|c|c|}
    \hline
        $N$ & 5 & 7 & 9 & 11 & 13 & 15\\\hline
        $\mathcal{N}^{(2,5)}(N,1)$ & 12 & 33 & 80 & 180 & 393 & 801 \\\hline
    \end{tabular}
        \caption{Dimension of spin $s=1$ sector truncated up to the level $N$}
        \label{tab5}
\end{table}

\begin{figure}[htbp]
\minipage{0.32\textwidth}
  \includegraphics[width=\linewidth]{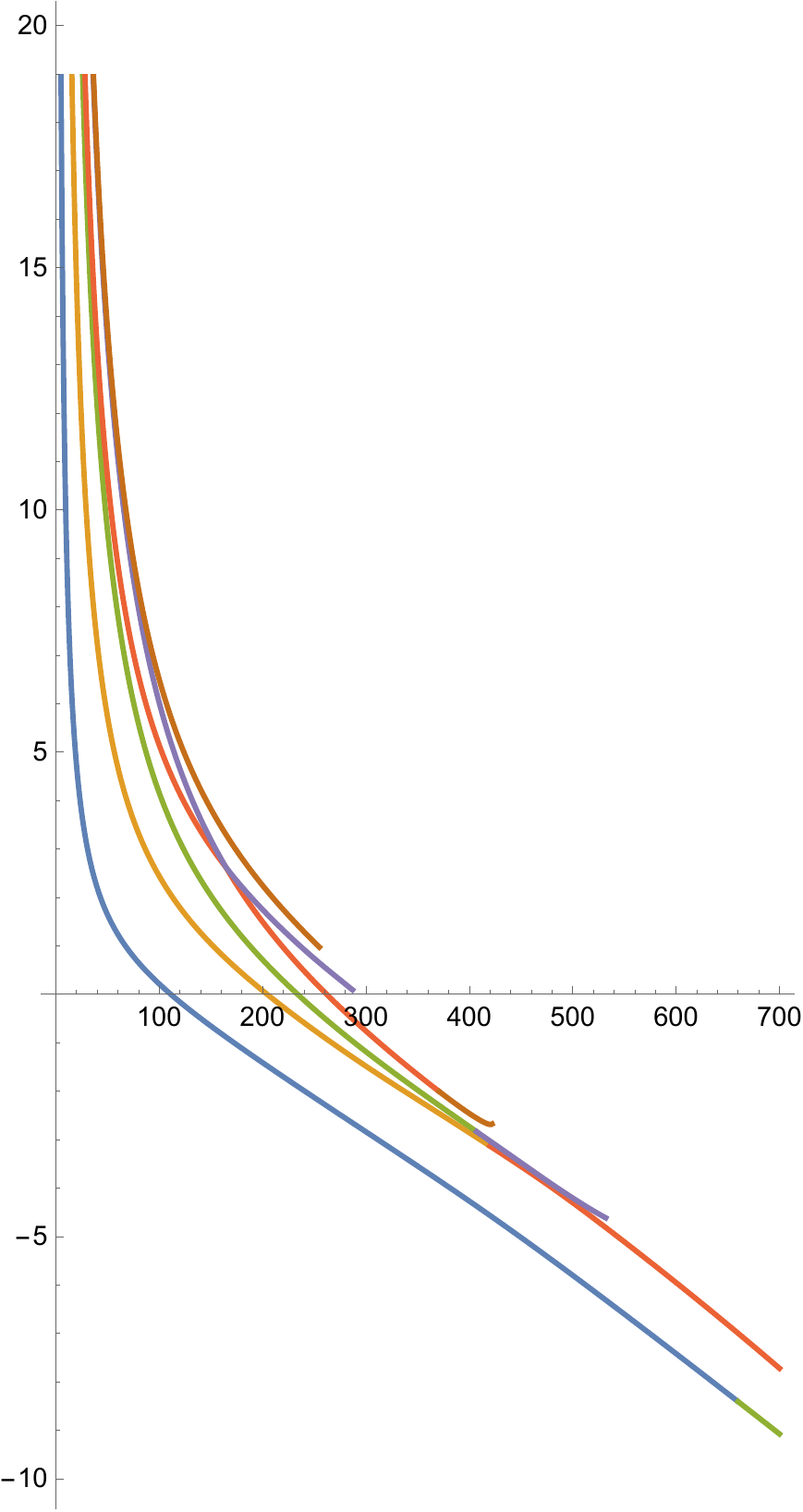}
  \caption{Energy levels $E_i(r)$ in spin $s=1$ sector. Truncation was made up to the level $N=5$.}\label{fig:4}
\endminipage\hfill
\minipage{0.32\textwidth}
  \includegraphics[width=\linewidth]{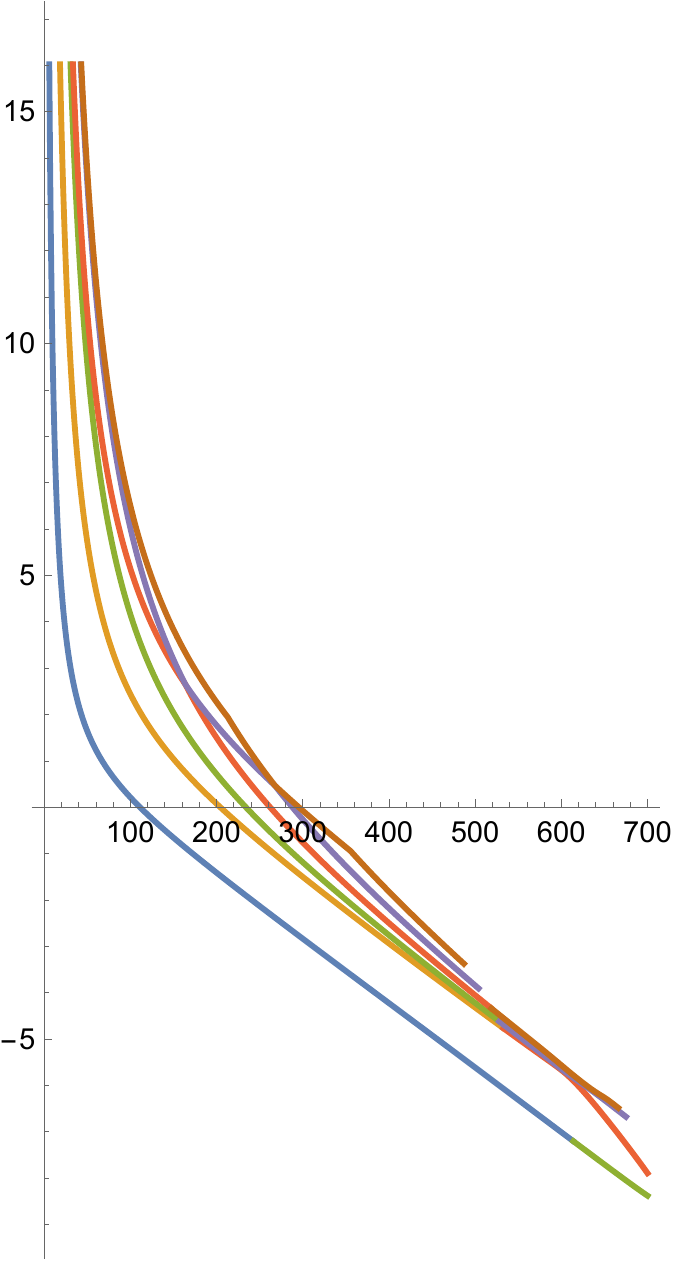}
  \caption{Energy levels $E_i(r)$ in spin $s=1$ sector. Truncation was made up to the level\\ $N=11$.}\label{fig:5}
\endminipage\hfill
\minipage{0.32\textwidth}%
  \includegraphics[width=\linewidth]{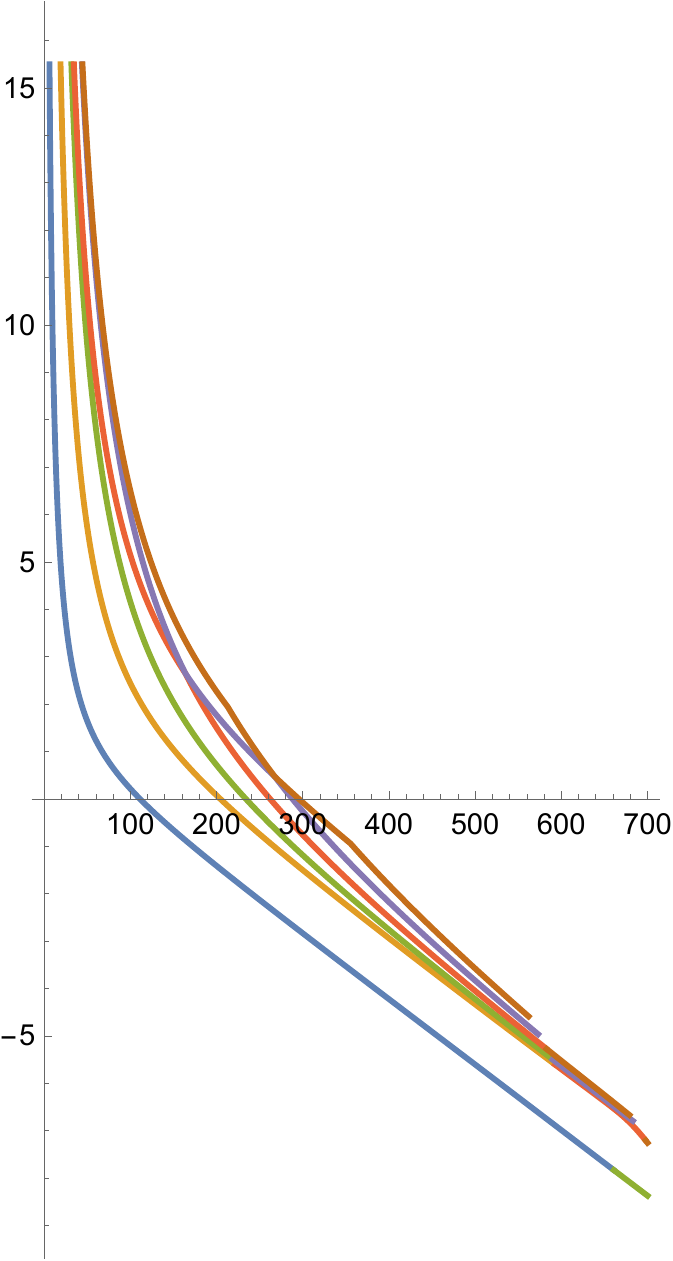}
  \caption{Energy levels $E_i(r)$ in spin $s=1$ sector. Truncation was made up to the level\\ $N=15$.}\label{fig:6}
\endminipage
\end{figure}

Using the TCSA algorithm, we can calculate the matrix elements of Hamiltonian $H$ for spin $s=1$ and different truncation levels $N$. The first 6 energy levels of $H$ against the scaling length $r$ for $N=5,11,15$ are plotted in Fig. \ref{fig:4}, \ref{fig:5}, \ref{fig:6}. The Fig. \ref{fig:4} with $N=5$ is analogous to Fig. 4 in \cite{Yurov:1989yu}.

The first 6 levels of the spectrum at large $r\rightarrow\infty$: the one-particle state $\partial\phi$, three two-particle states $\partial^2\bar{\partial}\phi$, $\partial^3\Bar{\partial}^2\phi$ and some linear combination of states $\bar{\partial}^3\phi_2$ \footnote{$\phi_2$ is a quasiprimary field at level 4 of the Verma module $\mathcal{V}_\phi$ in (\ref{phi2}).} and $\partial^4\bar{\partial}^3\phi$, the rest linear combination and $\bar{T}\partial T$ are two three-particle states.
\newpage
\subsubsection{Spin \texorpdfstring{$s=2$}{s=2} sector}
Consider Hilbert space with states of spin $s=2$. Size of matrix $H$ depending on truncation level $N$ is calculated using (\ref{eq11}) and represented in the Table \ref{tab6}.
\begin{table}[htbp]
        \centering
        \begin{tabular}{|c|c|c|c|c|c|c|}
    \hline
        $N$ & 5 & 7 & 9 & 11 & 13 & 15\\\hline
        $\mathcal{N}^{(2,5)}(N,2)$ & 9 & 25 & 64 & 147 & 319 & 669 \\\hline
    \end{tabular}
        \caption{Dimension of spin $s=2$ sector truncated up to the level $N$}
        \label{tab6}
\end{table}

\begin{figure}[t!]
\minipage{0.32\textwidth}
  \includegraphics[width=\linewidth]{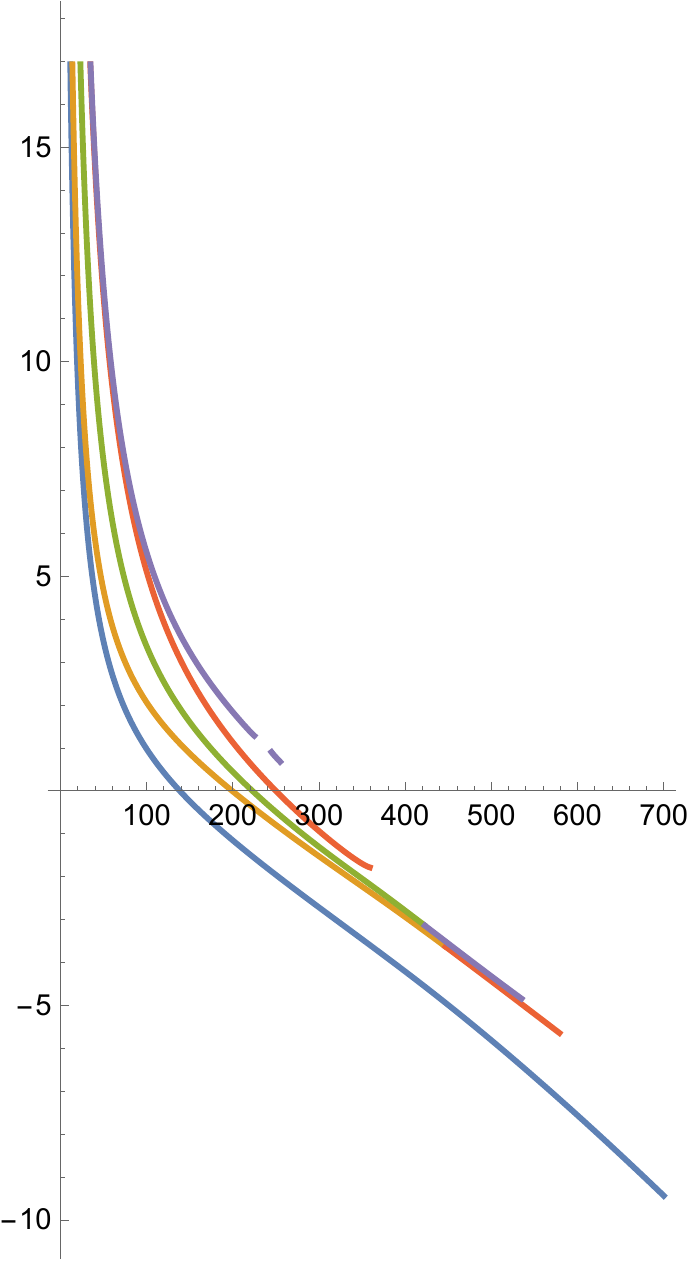}
  \caption{Energy levels $E_i(r)$ in spin $s=2$ sector. Truncation was made up to the level $N=5$.}\label{fig:7}
\endminipage\hfill
\minipage{0.32\textwidth}
  \includegraphics[width=\linewidth]{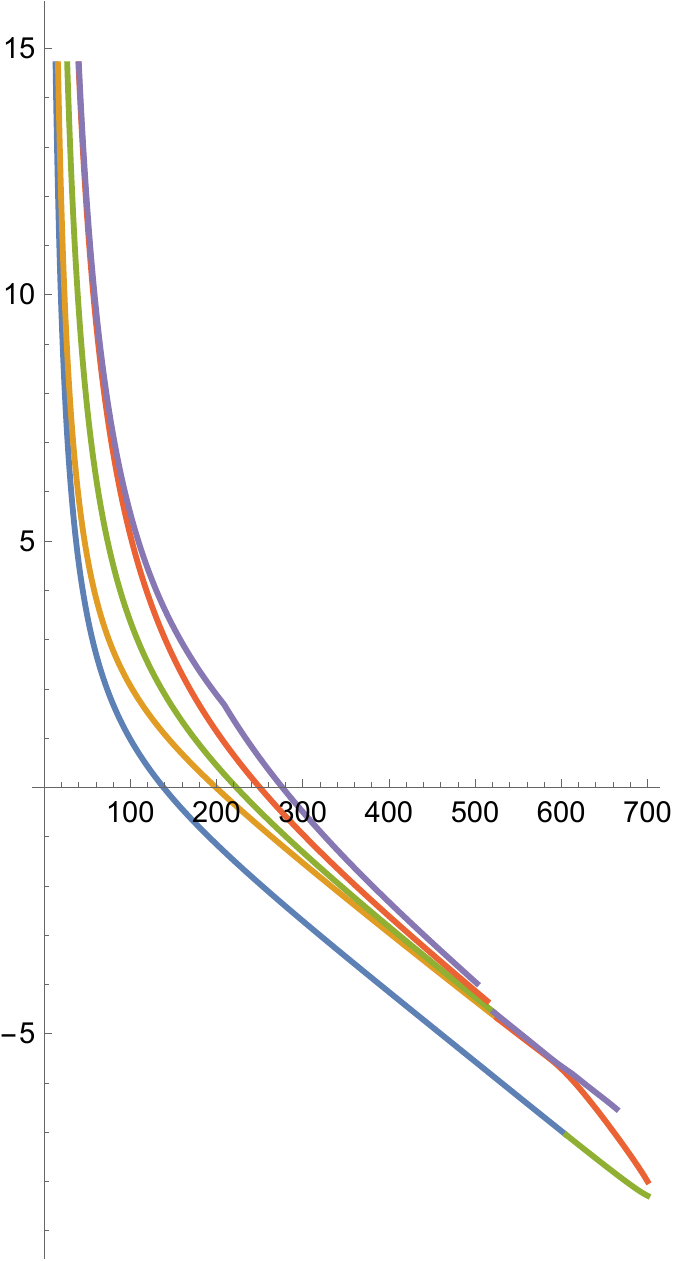}
  \caption{Energy levels $E_i(r)$ in spin $s=2$ sector. Truncation was made up to the level\\ $N=11$.}\label{fig:8}
\endminipage\hfill
\minipage{0.32\textwidth}%
  \includegraphics[width=\linewidth]{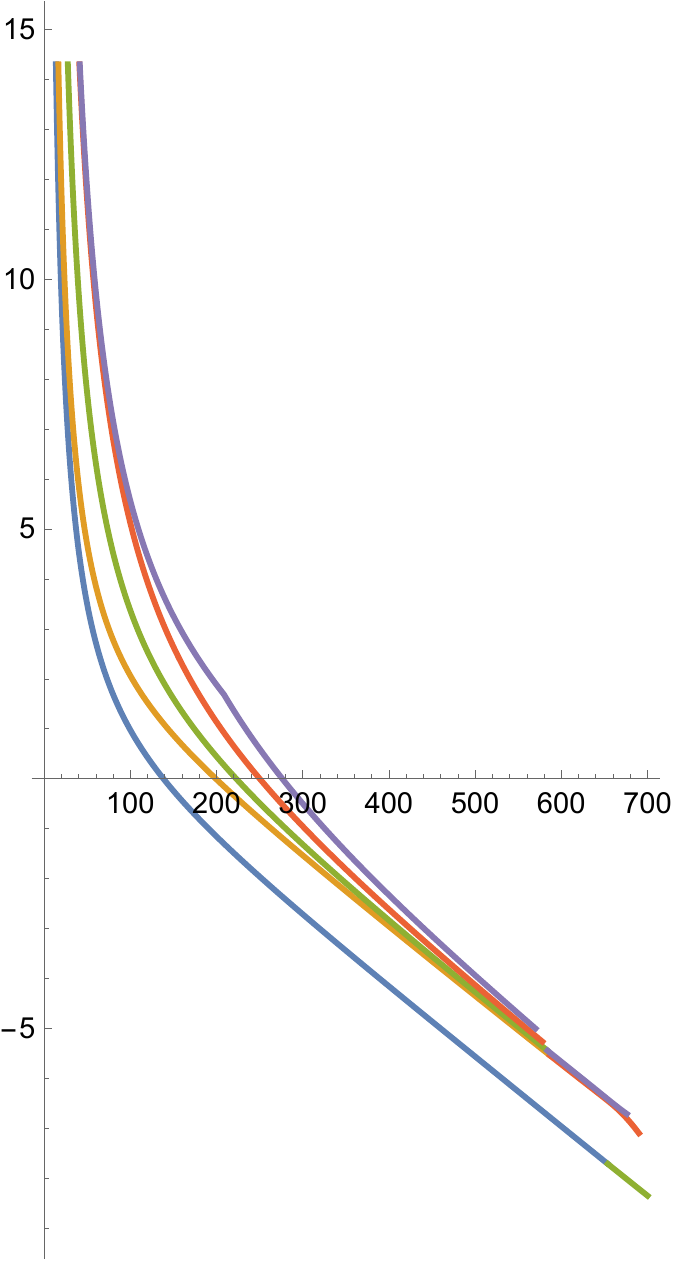}
  \caption{Energy levels $E_i(r)$ in spin $s=2$ sector. Truncation was made up to the level\\ $N=15$.}\label{fig:9}
\endminipage
\end{figure}

Using the TCSA algorithm, we can calculate the matrix elements of the Hamiltonian $H$ for spin $s=0$ and different truncation levels $N$. The first 5 energy levels of $H$ against the scaling length $r$ for $N=5,11,15$ are plotted in Fig. \ref{fig:7}, \ref{fig:8}, \ref{fig:9}. The Fig. \ref{fig:7} with $N=5$ is analogous to Fig. 5 in \cite{Yurov:1989yu}.
The first 5 levels of the spectrum at large $r\rightarrow\infty$: the one-particle state $\partial^2\phi$, three two-particle states $T$, $\partial^3\bar{\partial}\phi$ and some linear combination of states $\bar{\partial}^2\phi_2$ and $\partial^4\bar{\partial}^2\phi$, the rest linear combination and is a three-particle state.

\newpage
\subsection{\texorpdfstring{$M(3,10)$}{M(3,10)} perturbed by \texorpdfstring{$\phi_{1,3}$}{}}\label{TCSA1}
Let us perturb $M(3,10)$ by the relevant operator $\phi_{1,3}=\phi\otimes\phi$. Corresponding matrix element:
\begin{equation}
    \bra{\mathcal{O}_\gamma}\otimes\bra{\mathcal{O}_\delta}\phi\otimes\phi\ket{\mathcal{O}_\alpha}\otimes\ket{\mathcal{O}_\beta}=\bra{\mathcal{O}_\gamma}\phi\ket{\mathcal{O}_\alpha}\bra{\mathcal{O}_\delta}\phi\ket{\mathcal{O}_\beta}\,,
\end{equation}
where $\ket{\mathcal{O}_{\alpha,\beta,\gamma,\delta}}$ are derivatives of quasi-primary fields $\ket{h_{\alpha,\beta,\gamma,\delta}}$.

\begin{figure}[htbp]
\minipage{0.32\textwidth}
  \includegraphics[width=\linewidth]{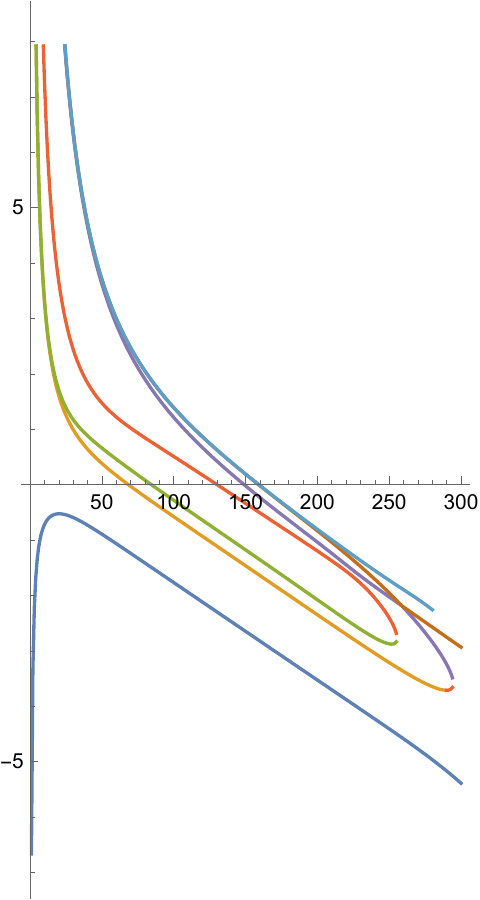}
  \caption{Energy levels $E_i(r)$ for $\phi_{1,3}$-perturbation. \\Truncation was made up to the energy $\mathcal{E}=5$.}\label{fig:10}
\endminipage\hfill
\minipage{0.32\textwidth}
  \includegraphics[width=\linewidth]{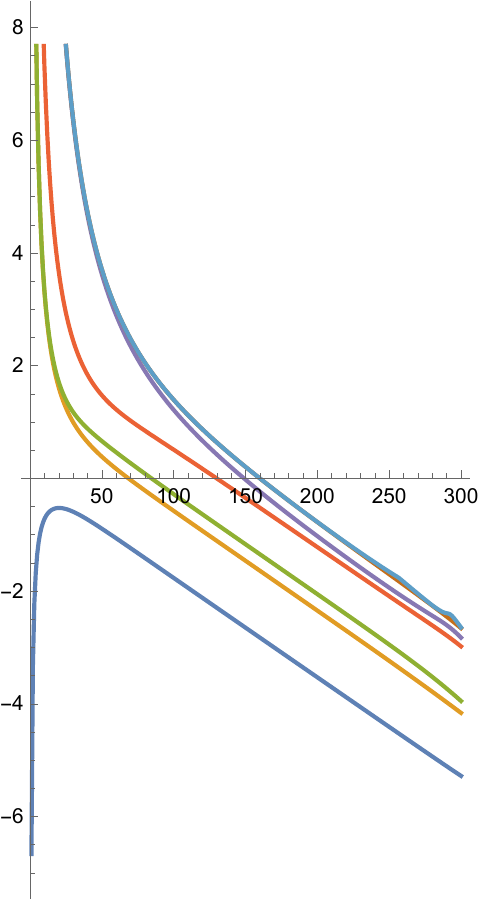}
  \caption{Energy levels $E_i(r)$ for $\phi_{1,3}$-perturbation. \\Truncation was made up to the energy $\mathcal{E}=9$.}\label{fig:11}
\endminipage\hfill
\minipage{0.32\textwidth}%
  \includegraphics[width=\linewidth]{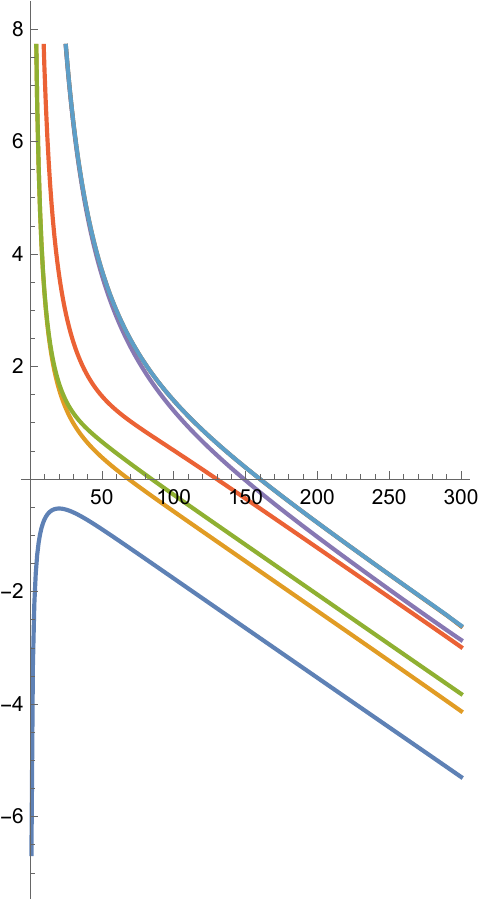}
  \caption{Energy levels $E_i(r)$ for $\phi_{1,3}$-perturbation. \\Truncation was made up to the energy $\mathcal{E}=13$.}\label{fig:12}
\endminipage
\end{figure}

Using the TCSA algorithm, we can calculate the matrix elements of the Hamiltonian $H$ for spin $s=0$ and different truncation energies $\mathcal{E}$. The first 7 energy levels of $H$ against the scaling length $r$ for $\mathcal{E}=5,9,13$ are plotted in Fig. \ref{fig:10}, \ref{fig:11}, \ref{fig:12}.

\begin{figure}[htbp]
    \centering
    \includegraphics[scale=0.44]{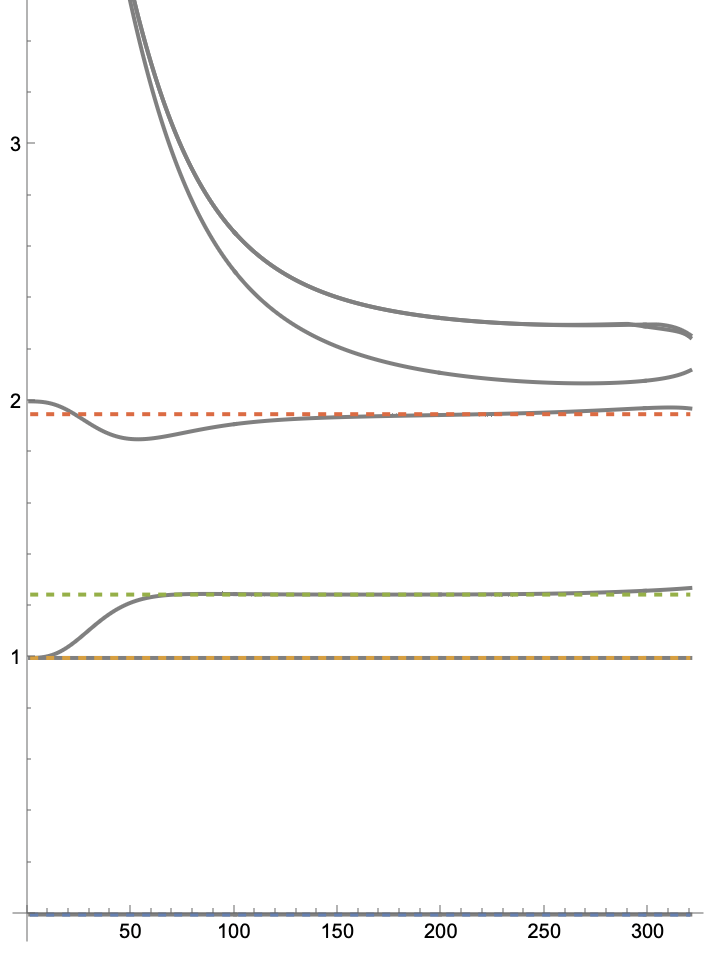}
    \caption{$\frac{E_i-E_0}{E_1-E_0}(r)$ for $\phi_{1,3}$-perturbation. Truncation was made up to the energy $\mathcal{E}=13$. Dashed blue corresponds to vacuum, yellow to kink $m_{\text{kink}}$, green to the first breather $m_1$, red to the second breather $m_2$.}\label{fig:14}
\end{figure}

The plot of $\frac{E_i-E_0}{E_1-E_0}$, where $i\in\{1,...,7\}$ is the corresponding energy level, against the scaling length $r$ for $\mathcal{E}=13$ is depicted in Fig. \ref{fig:14}. Together with the numerical data, theoretical predictions are shown (from subsection \ref{phi13}): vacuum $m=0$, kink $m_{\text{kink}}$ and two breathers with masses (\ref{conj}). As we can see from the plot, TCSA results support the conjecture (\ref{conj}).


\newpage
\subsection{\texorpdfstring{$M(3,10)$}{M(3,10)} perturbed by \texorpdfstring{$i\phi^+_{1,5}$}{}}\label{TCSA2}
Let us perturb $M(3,10)$ by $\mathbb{Z}_2$-even operator $i\phi^+_{1,5}=I\otimes i\phi+i\phi\otimes I$. Corresponding matrix element:
\begin{equation}
    \bra{\mathcal{O}_\gamma}\otimes\bra{\mathcal{O}_\delta}(I\otimes\phi+\phi\otimes I)\ket{\mathcal{O}_\alpha}\otimes\ket{\mathcal{O}_\beta}=\bra{\mathcal{O}_\gamma}I\ket{\mathcal{O}_\alpha}\bra{\mathcal{O}_\delta}\phi\ket{\mathcal{O}_\beta}+\bra{\mathcal{O}_\gamma}\phi\ket{\mathcal{O}_\alpha}\bra{\mathcal{O}_\delta}I\ket{\mathcal{O}_\beta}\,,
\end{equation}
where $\ket{\mathcal{O}_{\alpha,\beta,\gamma,\delta}}$ are derivatives of quasi-primary fields $\ket{h_{\alpha,\beta,\gamma,\delta}}$.

\begin{figure}[htbp]
\minipage{0.32\textwidth}
  \includegraphics[width=\linewidth]{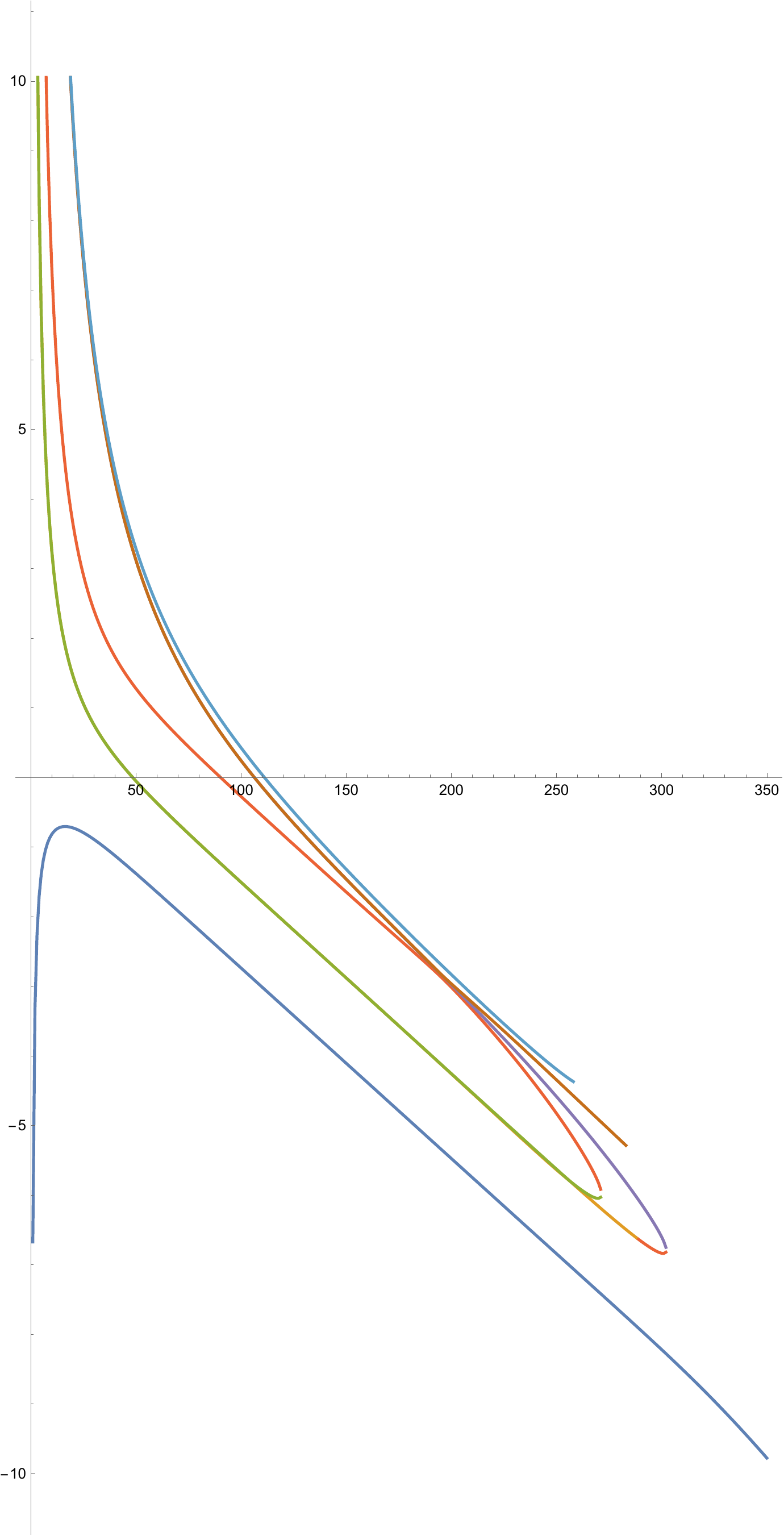}
  \caption{Energy levels $E_i(r)$ for $i\phi^+_{1,5}$-perturbation. \\Truncation was made up to the energy $\mathcal{E}=5$.}\label{fig:15}
\endminipage\hfill
\minipage{0.32\textwidth}
  \includegraphics[width=\linewidth]{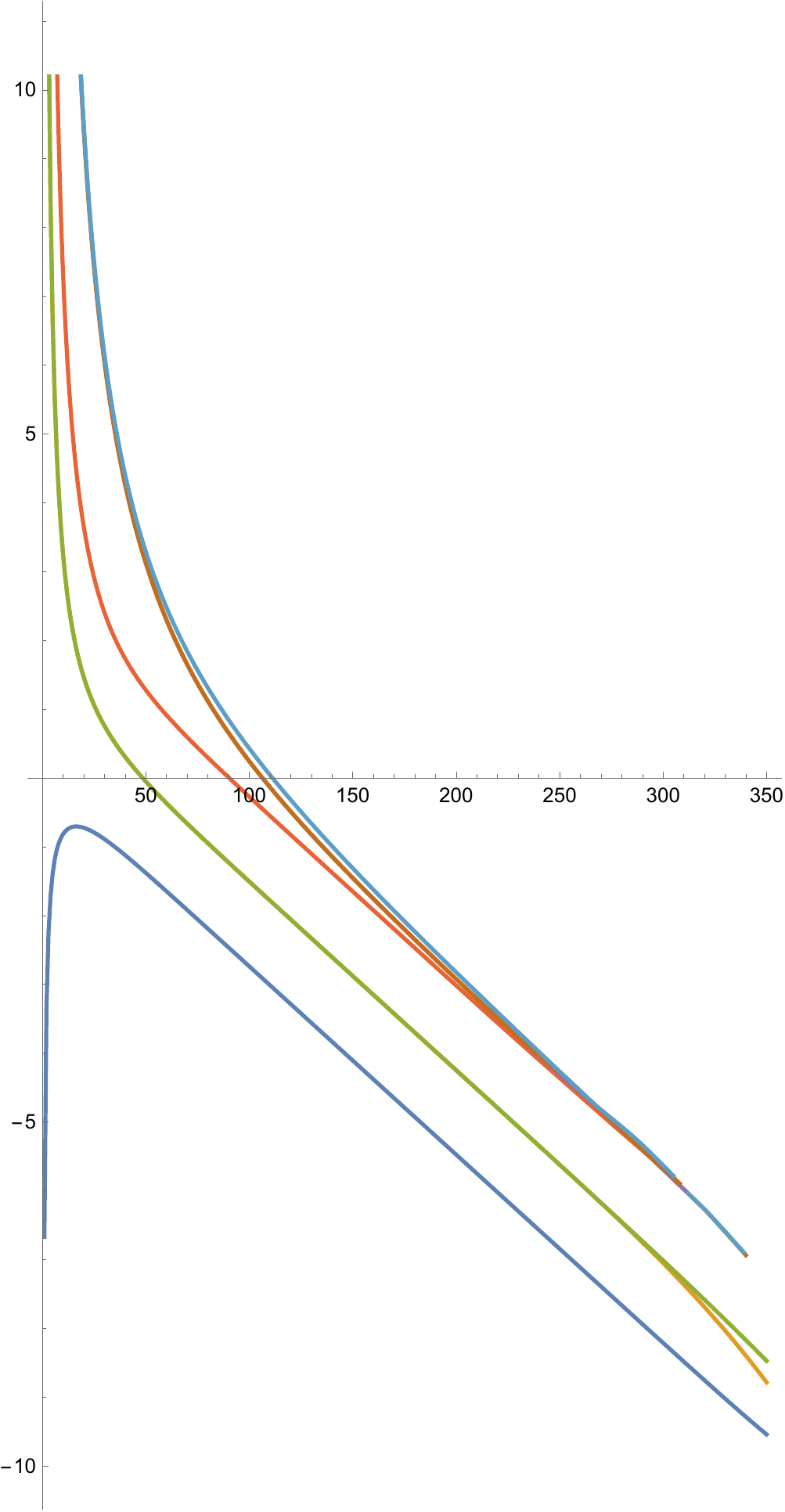}
  \caption{Energy levels $E_i(r)$ for $i\phi^+_{1,5}$-perturbation. \\Truncation was made up to the energy $\mathcal{E}=9$.}\label{fig:16}
\endminipage\hfill
\minipage{0.32\textwidth}%
  \includegraphics[width=\linewidth]{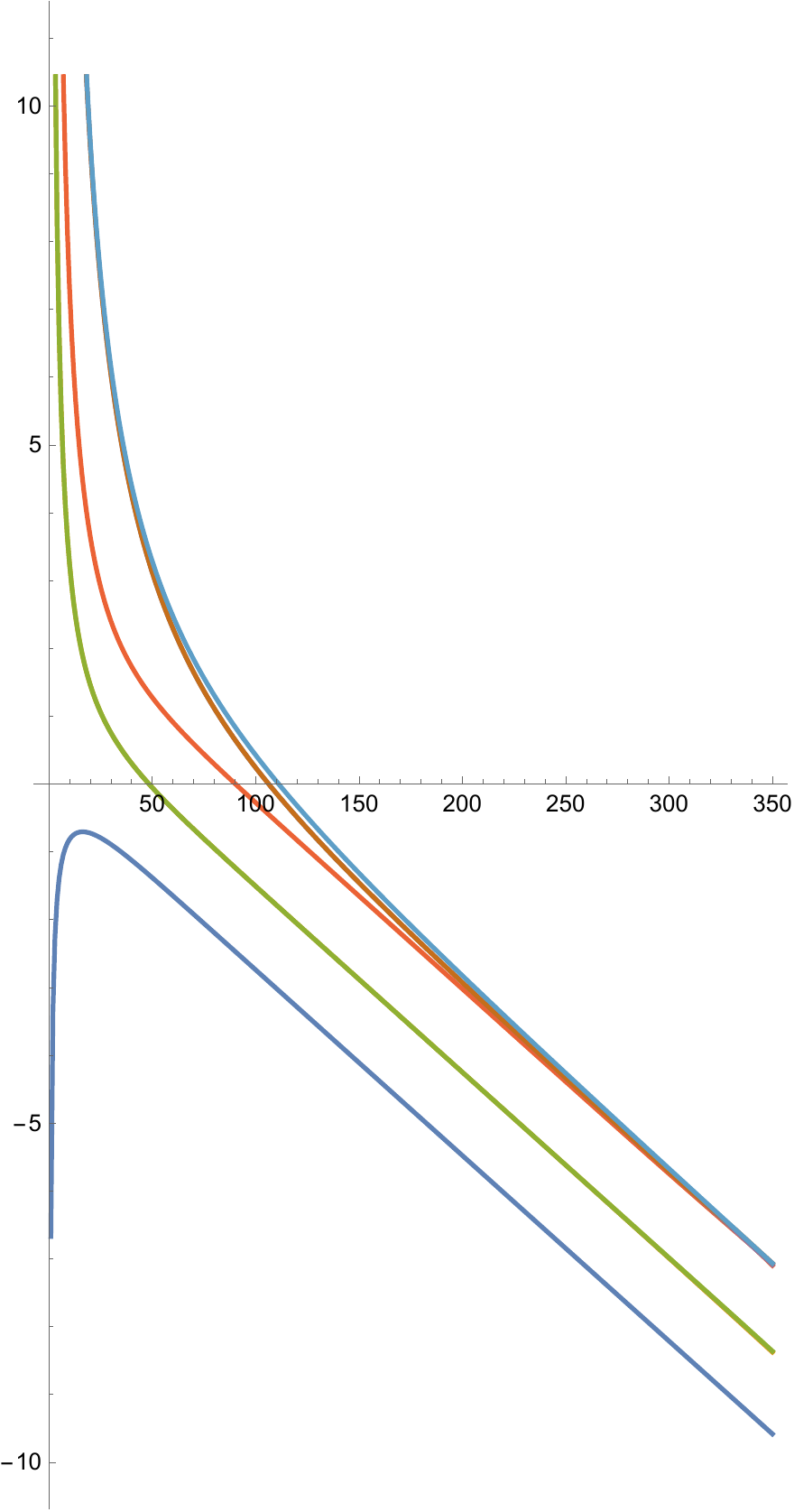}
  \caption{Energy levels $E_i(r)$ for $i\phi^+_{1,5}$-perturbation. \\Truncation was made up to the energy $\mathcal{E}=13$.}\label{fig:17}
\endminipage
\end{figure}

Using the TCSA algorithm, we can calculate the matrix elements of the Hamiltonian $H$ for spin $s=0$ and different truncation energies $\mathcal{E}$. The first 7 energy levels of $H$ against the scaling length $r$ for $\mathcal{E}=5,9,13$ are plotted in Fig. \ref{fig:15}, \ref{fig:16}, \ref{fig:17}.

\begin{figure}[htbp]
    \centering
    \includegraphics[scale=0.35]{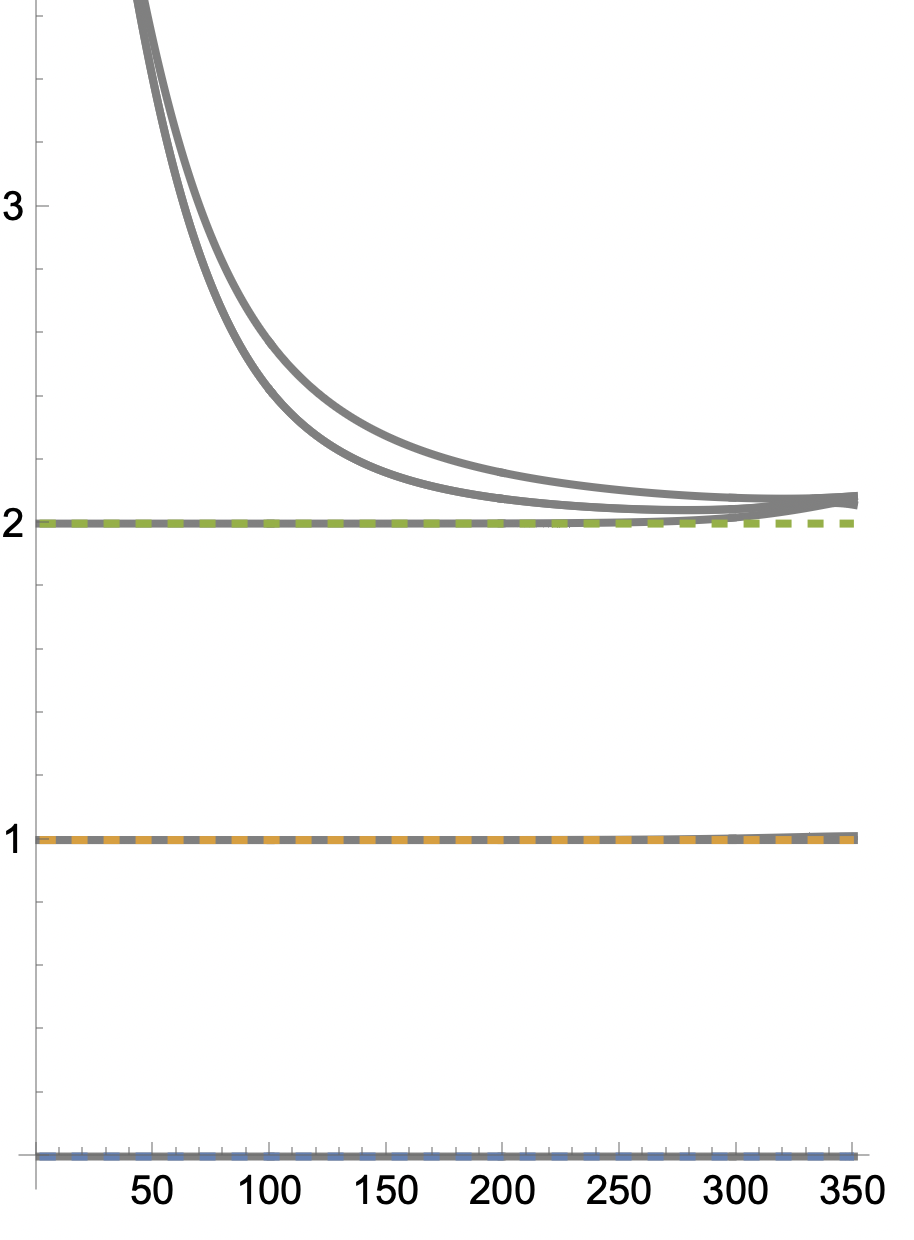}
    \caption{$\frac{E_i-E_0}{E_1-E_0}(r)$ for $i\phi^+_{1,5}$-perturbation. Truncation was made up to the energy $\mathcal{E}=13$. Dashed blue corresponds to vacuum, yellow to two one-breather states $m$, green to two-breather states $2m$.}\label{fig:18}
\end{figure}

The plot of $\frac{E_i-E_0}{E_1-E_0}$, where $i\in\{1,...,7\}$ is the corresponding energy level, against the scaling length $r$ for $\mathcal{E}=13$ is depicted in Fig. \ref{fig:18}. Together with the numerical data, theoretical predictions are shown (from Subsection \ref{phi15}).

\section{Discussion}
In this paper, we have applied TCSA for SLYM and $M(3,10)$ perturbed by relevant fields $\phi_{1,3}$ and $i\phi^+_{1,5}$. We have found quasi-primary fields in $M(2,5)$ and reproduced classic results using TCSA up to the level $N=15$. Using the theory of reduced sine-Gordon theory, we have made the conjecture about the spectrum of $M(3,10)+\phi_{1,3}$ (\ref{conj}), which was in agreement with TCSA. Also, using TCSA, we have checked the well-known spectrum of $M(3,10)+i\phi^+_{1,5}$.

In the future, it may be interesting to obtain the spectrum of other tensor products of minimal model \cite{Quella:2006de}: $E_6$ series version of $M(5,12)=M(2,5)\otimes M(3,4)$ and $E_8$ series version of $M(7,30)=M(2,5)\otimes M(2,7)$ perturbed by relevant fields. Also, it is interesting to see the RG flow  $M(3,10)+\phi_{1,7}\rightarrow M(3,8)$ \cite{Fei:2014xta,Klebanov:2022syt,Katsevich:2024jgq} using TCSA.

\acknowledgments

The significant part of this work was done in my Bachelor's MIPT thesis (2023) under the supervision of Alexei Litvinov, who inspired me to work on this project. The author is grateful to acknowledge very helpful discussions with my current PhD advisor Igor Klebanov, as well as with Michael Lashkevich, Zimo Sun and Grigory Tarnopolsky. 
This work was supported in part by the US National Science Foundation Grant No. PHY-2209997 and by the Simons Foundation Grant No. 917464.

\appendix
\section{List of quasi-primary fields in \texorpdfstring{$M(2,5)$}{M(2,5)}}\label{qfields}
Quasi-primary fields for Verma module $\mathcal{V}_{1,1}$ up to the level 15:
\begin{equation}
    \ket{I_1}=\ket{I}\,;
\end{equation}
\begin{equation}
    \ket{I_2}=i\sqrt{\frac{5}{11}}L_{-2}\ket{I}\,;
\end{equation}
\begin{equation}
    \ket{I_3}=\sqrt{\frac{162}{11935}}\left(L_{-6}-\frac{35}{36}L^2_{-3}\right)\ket{I}\,;
\end{equation}
\begin{equation}
    \ket{I_4}=\sqrt{\frac{25}{70356}}\left(L_{-8}-3L_{-5}L_{-3}+\frac{3}{5}L_{-4}^2\right)\ket{I}\,;
\end{equation}
\begin{equation}
    \ket{I_5}=\sqrt{\frac{18}{63767}}\left(L_{-10}-\frac{11}{12}L_{-7}L_{-3}+\frac{11}{15}L_{-6}L_{-4}-\frac{11}{15}L_{-5}^2\right)\ket{I}\,;
\end{equation}
\begin{equation}
\begin{aligned}
    \ket{I_6}&=\frac{25}{2}\sqrt{\frac{35}{4559131434}} \left(-\frac{39}{10} L_{-9}L_{-3}+\frac{117}{25} L_{-8}L_{-4}-\right.\\
    &\left.-\frac{481}{50} L_{-7}L_{-5}+\frac{767}{175}L^2_{-6}+L_{-12}\right)\ket{I}\,;
\end{aligned}
\end{equation}
\begin{equation}
\begin{aligned}
    \ket{I_7}&=i\frac{153671}{10}\sqrt{\frac{3}{140300872910}} \left(-\frac{430729}{922026}L_{-9}L_{-3}-\frac{110065}{461013}L_{-8}L_{-4}-\right.\\
    &\left.-\frac{264043}{922026}L_{-7}L_{-5}-\frac{33601}{65859} L^2_{-6}+\frac{164510}{461013}L_{-6}L_{-3}^2+L_{-12}\right)\ket{I}\,;
\end{aligned}
\end{equation}
\begin{equation}
\begin{aligned}
    \ket{I_8}&=18\sqrt{\frac{5}{173998979}} \left(-\frac{7}{18} L_{-11}L_{-3}+\frac{28}{45} L_{-10}L_{-4}-\right.\\
    &\left.-\frac{7}{5} L_{-9}L_{-5}+\frac{16}{9}L_{-8}L_{-6}-\frac{79}{72}L_{-7}^2+L_{-14}\right)\ket{I}\,;
\end{aligned}
\end{equation}
\begin{equation}
\begin{aligned}
    \ket{I_9}=i\frac{247416}{25}\sqrt{\frac{2}{25664991605}} \left(-\frac{20987}{76128}L_{-11}L_{-3}-\frac{7301}{30927}L_{-10}L_{-4}-\frac{31321}{164944}L_{-9}L_{-5}-\right.\\\left.-\frac{58289}{247416}L_{-8}L_{-6}-\frac{71525}{247416}L^2_{-7}+\frac{25855}{989664}L_{-8}L^2_{-3}+\frac{77565}{329888}L_{-7}L_{-4}L_{-3}+L_{-14}\right)\ket{I}\,;
\end{aligned}
\end{equation}
\begin{equation}
\begin{aligned}
    \ket{I_{10}}&=i\frac{1153}{22} \sqrt{\frac{3}{535990}}\left(-\frac{365}{1153}L_{-12}L_{-3}-\frac{179}{1153}L_{-11}L_{-4}-\frac{252}{1153}L_{-10}L_{-5}-\right.\\
    &\left.-\frac{306}{1153}L_{-9}L_{-6}-\frac{1160}{3459} L_{-8}L_{-7}+\frac{165}{2306}L_{-9}L^2_{-3}+\frac{275}{3459}L_{-8}L_{-4}L_{-3}+\right.\\
    &\left.+\frac{275}{2306}L_{-7}L_{-5}L_{-3}+L_{-15}\right)\ket{I}\,.
\end{aligned}
\end{equation}
Quasi-primary fields for the Verma module $\mathcal{V}_{1,2}$ up to the level 15:
\begin{equation}
    \ket{\phi_1}=\ket{\phi}\,;
\end{equation}
\begin{equation}\label{phi2}
    \ket{\phi_2}=\sqrt{\frac{5}{1482}}\left(L_{-4}-\frac{25}{2}L_{-3}L_{-1}\right)\ket{\phi}\,;
\end{equation}
\begin{equation}
    \ket{\phi_3}=62\sqrt{\frac{5}{1634817}}\left(-\frac{35}{31}L_{-5}L_{-1}-\frac{525}{124} L_{-3}L_{-2}L_{-1}+L_{-6}\right)\ket{\phi}\,;
\end{equation}
\begin{equation}
    \ket{\phi_4}=\frac{25}{\sqrt{236698}}\left(-\frac{5}{2} L_{-7}L_{-1}-\frac{6}{5}L_{-4}^2+5 L_{-4}L_{-3}L_{-1}+L_{-8}\right)\ket{\phi}\,;
\end{equation}
\begin{equation}
\begin{aligned}
    \ket{\phi_5}=i\frac{7}{2}\sqrt{\frac{5}{4018443}}\left(-\frac{25}{2}L_{-8}L_{-1}-\frac{5}{7}L_{-5}L_{-4}-\right.\\\left.-\frac{125}{7} L_{-5}L_{-3}L_{-1}+\frac{225}{14}L_{-4}^2L_{-1}+L_{-9}\right)\ket{\phi}\,;
\end{aligned}
\end{equation}
\begin{equation}
\begin{aligned}
    \ket{\phi_6}=47\sqrt{\frac{2}{4950203}}\left(-\frac{605}{188} L_{-9}L_{-1}-\frac{385}{282} L_{-6}L_{-4}-\frac{55}{564} L_{-5}^2+\right.\\
    \left.+\frac{1375}{564}L_{-6}L_{-3}L_{-1}+\frac{1375}{564}L_{-5}L_{-4}L_{-1}+L_{-10}\right)\ket{\phi}\,;
\end{aligned}
\end{equation}
\begin{equation}
\begin{aligned}
    \ket{\phi_7}=\frac{1}{11} i \sqrt{\frac{5}{13702}}\left(-10 L_{-10}L_{-1}-\frac{15}{7} L_{-7}L_{-4}+\frac{100}{7}L_{-7}L_{-3}L_{-1}-\right.\\\left.-\frac{125}{7} L_{-6}L_{-4}L_{-1}+\frac{25}{2}L_{-5}^2L_{-1}+L_{-11}\right)\ket{\phi}\,;
\end{aligned}
\end{equation}
\begin{equation}
\begin{aligned}
    \ket{\phi_8}=40\sqrt{\frac{105}{698596813}}\left(-\frac{35}{16} L_{-11}L_{-1}-\frac{113}{48} L_{-8}L_{-4}-\frac{41}{56} L_{-6}^2+\frac{95}{96}L_{-8}L_{-3}L_{-1}+\right.\\\left.+\frac{55}{32}L_{-7}L_{-4}L_{-1}+\frac{5}{8} L_{-6}L_{-5}L_{-1}+\frac{7}{24} L_{-4}^3+L_{-12}\right)\ket{\phi}\,;
\end{aligned}
\end{equation}
\begin{equation}
\begin{aligned}
    \ket{\phi_9}&=i\frac{6172859615}{14} \sqrt{\frac{3}{46573185837240060252974}}\left(-\frac{34566275425}{11111147307}L_{-11}L_{-1}+\right.\\
    &\left.+\frac{144867739664}{11111147307}L_{-8}L_{-4}+\frac{698596813}{854703639}L_{-7}L_{-5}+\frac{23089835417}{11111147307}L^2_{-6}+\right.\\
    &\left.+\frac{10231315925}{11111147307}L_{-8}L_{-3}L_{-1}-\frac{17975971225}{3703715769}L_{-7}L_{-4}L_{-1}+\right.\\
    &\left.+\frac{4701214525}{1234571923}L_{-6}L_{-5}L_{-1}-\frac{25663691840}{11111147307}L_{-4}^3+L_{-12}\right)\ket{\phi}\,;
\end{aligned}
\end{equation}
\begin{equation}
\begin{aligned}
    \ket{\phi_{10}}=i\frac{200}{3}\sqrt{\frac{2}{1533403333}} \left(-\frac{7}{16}L_{-12}L_{-1}-\frac{63}{80}L_{-9}L_{-4}+\frac{7}{40}L_{-8}L_{-5}-\frac{41}{80}L_{-7}L_{-6}+\right.\\\left.+\frac{105}{32}L_{-9}L_{-3}L_{-1}-\frac{63}{16}L_{-8}L_{-4}L_{-1}+\frac{259}{32}L_{-7}L_{-5}L_{-1}-\frac{59}{16}L_{-6}^2L_{-1}+L_{-13}\right)\ket{\phi}\,;
\end{aligned}
\end{equation}
\begin{equation}
\begin{aligned}
    \ket{\phi_{11}}&=i\frac{3872}{3}\sqrt{\frac{5}{79528560713}} \left(\frac{925}{704}L_{-13}L_{-1}-\frac{13195}{7744}L_{-10}L_{-4}+\frac{35}{64}L_{-9}L_{-5}-\right.\\
    &\left.-\frac{3175}{1408}L_{-8}L_{-6}+\frac{25}{1408}L_{-7}^2-\frac{875}{1936}L_{-10}L_{-3}L_{-1}+\frac{175}{352}L_{-9}L_{-4}L_{-1}-\right.\\
    &\left.-\frac{875}{704}L_{-8}L_{-5}L_{-1}+\frac{1225}{1408}L_{-6}L_{-4}^2-\frac{175}{704}L_{-5}^2L_{-4}+L_{-14}\right)\ket{\phi}\,;
\end{aligned}
\end{equation}
\begin{equation}
\begin{aligned}
    \ket{\phi_{12}}=\frac{373623381}{2}\sqrt{\frac{5}{1312291077602220331861}}\left(-\frac{4516340575}{8966961144}L_{-13}L_{-1}-\right.\\
    \left.-\frac{3871710367}{2988987048}L_{-10}L_{-4}+\frac{4020247525}{8966961144}L_{-9}L_{-5}-\frac{33119831795}{17933922288}L_{-8}L_{-6}+\right.\\
    \left.+\frac{31951075}{221406448}L_{-7}^2+\frac{104326075}{1120870143}L_{-10}L_{-3}L_{-1}+\frac{2993011105}{4483480572}L_{-9}L_{-4}L_{-1}-\right.\\
    \left.-\frac{3458236075}{8966961144}L_{-8}L_{-5}L_{-1}+\frac{821915675}{2241740286}L_{-7}L_{-6}L_{-1}+\right.\\
    \left.+\frac{3148086095}{5977974096}L_{-6}L_{-4}^2-\frac{449726585}{2988987048}L^2_{-5}L_{-4}+L_{-14}\right)\ket{\phi}\,;
\end{aligned}
\end{equation}
\begin{equation}
\begin{aligned}
    \ket{\phi_{13}}&=i\frac{165}{2 \sqrt{4835683438}}\left(-\frac{60}{11}L_{-14}L_{-1}-\frac{61}{99} L_{-11}L_{-4}+\frac{10}{33} L_{-10}L_{-5}-\frac{74}{99}L_{-9}L_{-6}-\right.\\
    &\left.-\frac{5}{198}L_{-8}L_{-7}+\frac{200}{99} L_{-11}L_{-3}L_{-1}-\frac{85}{33} L_{-10}L_{-4}L_{-1}+\frac{655}{99} L_{-9}L_{-5}L_{-1}-\right.\\
    &\left.-\frac{775}{99} L_{-8}L_{-6}L_{-1}+\frac{225}{44}L_{-7}^2L_{-1}+L_{-15}\right)\ket{\phi}\,;
\end{aligned}
\end{equation}
\begin{equation}
\begin{aligned}
    \ket{\phi_{14}}&=i\frac{587458735}{7392} \sqrt{\frac{19}{33706398798538143}} \left(\frac{2636446870}{2232343193}L_{-14}L_{-1}-\frac{1351061546}{2232343193}L_{-11}L_{-4}+\right.\\
    &\left.+\frac{1033250602}{2232343193}L_{-10}L_{-5}+\frac{3954958994}{2232343193}L_{-9}L_{-6}-\frac{5570049565}{2232343193}L_{-8}L_{-7}-\right.\\&\left.-\frac{1171135050}{2232343193}L_{-11}L_{-3}L_{-1}+\frac{573366100}{2232343193}L_{-10}L_{-4}L_{-1}-\frac{2915636200}{2232343193}L_{-9}L_{-5}L_{-1}-\right.\\
    &\left.-\frac{2689095950}{2232343193}L_{-8}L_{-6}L_{-1}+\frac{1798975400}{2232343193}L_{-7}^2L_{-1}+\frac{3679324355}{2232343193}L_{-7}L_{-4}^2-\right.\\
    &\left.-\frac{3153706590}{2232343193}L_{-6}L_{-5}L_{-4}+\frac{1051235530}{6697029579}L_{-5}^3+L_{-15}\right)\ket{\phi}\,.
\end{aligned}
\end{equation}

\bibliographystyle{JHEP}
\bibliography{biblio.bib}

\providecommand{\href}[2]{#2}\begingroup\raggedright\begin{thebibliography}{10}

\bibitem{Belavin:1984vu}
A.A.~Belavin, A.M.~Polyakov and A.B.~Zamolodchikov, \emph{{Infinite Conformal Symmetry in Two-Dimensional Quantum Field Theory}}, \href{https://doi.org/10.1016/0550-3213(84)90052-X}{\emph{Nucl. Phys. B} {\bfseries 241} (1984) 333}.

\bibitem{zbMATH03661547}
V.G.~Kac, ``Highest weight representations of infinite-dimensional {Lie} algebras.'' Proc. int. {Congr}. {Math}., {Helsinki} 1978, {Vol}. 1, 299-304 (1980).

\bibitem{Xu:2022mmw}
H.-L.~Xu and A.~Zamolodchikov, \emph{{2D Ising Field Theory in a magnetic field: the Yang-Lee singularity}}, \href{https://doi.org/10.1007/JHEP08(2022)057}{\emph{JHEP} {\bfseries 08} (2022) 057} [\href{https://arxiv.org/abs/2203.11262}{{\ttfamily 2203.11262}}].

\bibitem{Xu:2023nke}
H.-L.~Xu and A.~Zamolodchikov, \emph{{Ising Field Theory in a magnetic field: $\varphi^3$ coupling at $T>T_c$}}, \href{https://doi.org/10.1007/JHEP08(2023)161}{\emph{JHEP} {\bfseries 08} (2023) 161} [\href{https://arxiv.org/abs/2304.07886}{{\ttfamily 2304.07886}}].

\bibitem{Xu:2024baz}
H.-L.~Xu, \emph{{On the analyticity of the lightest particle mass of Ising field theory in a magnetic field}},  \href{https://arxiv.org/abs/2405.09091}{{\ttfamily 2405.09091}}.

\bibitem{Nakayama:2022svf}
Y.~Nakayama and K.~Kikuchi, \emph{{The fate of non-supersymmetric Gross-Neveu-Yukawa fixed point in two dimensions}}, \href{https://doi.org/10.1007/JHEP03(2023)240}{\emph{JHEP} {\bfseries 03} (2023) 240} [\href{https://arxiv.org/abs/2212.06342}{{\ttfamily 2212.06342}}].

\bibitem{Nakayama:2024msv}
Y.~Nakayama and T.~Tanaka, \emph{{Infinitely many new renormalization group flows between Virasoro minimal models from non-invertible symmetries}}, \href{https://doi.org/10.1007/JHEP11(2024)137}{\emph{JHEP} {\bfseries 11} (2024) 137} [\href{https://arxiv.org/abs/2407.21353}{{\ttfamily 2407.21353}}].

\bibitem{Klebanov:2022syt}
I.R.~Klebanov, V.~Narovlansky, Z.~Sun and G.~Tarnopolsky, \emph{{Ginzburg-Landau description and emergent supersymmetry of the (3, 8) minimal model}}, \href{https://doi.org/10.1007/JHEP02(2023)066}{\emph{JHEP} {\bfseries 02} (2023) 066} [\href{https://arxiv.org/abs/2211.07029}{{\ttfamily 2211.07029}}].

\bibitem{Katsevich:2024jgq}
A.~Katsevich, I.R.~Klebanov and Z.~Sun, \emph{{Ginzburg-Landau description of a class of non-unitary minimal models}},  \href{https://arxiv.org/abs/2410.11714}{{\ttfamily 2410.11714}}.

\bibitem{Delouche:2023wsl}
O.~Delouche, J.~Elias~Miro and J.~Ingoldby, \emph{{Hamiltonian truncation crafted for UV-divergent QFTs}}, \href{https://doi.org/10.21468/SciPostPhys.16.4.105}{\emph{SciPost Phys.} {\bfseries 16} (2024) 105} [\href{https://arxiv.org/abs/2312.09221}{{\ttfamily 2312.09221}}].

\bibitem{Lencses:2022ira}
M.~Lencs\'es, A.~Miscioscia, G.~Mussardo and G.~Tak\'acs, \emph{{Multicriticality in Yang-Lee edge singularity}}, \href{https://doi.org/10.1007/JHEP02(2023)046}{\emph{JHEP} {\bfseries 02} (2023) 046} [\href{https://arxiv.org/abs/2211.01123}{{\ttfamily 2211.01123}}].

\bibitem{Lencses:2023evr}
M.~Lencs\'es, A.~Miscioscia, G.~Mussardo and G.~Tak\'acs, \emph{{$ \mathcal{PT} $ breaking and RG flows between multicritical Yang-Lee fixed points}}, \href{https://doi.org/10.1007/JHEP09(2023)052}{\emph{JHEP} {\bfseries 09} (2023) 052} [\href{https://arxiv.org/abs/2304.08522}{{\ttfamily 2304.08522}}].

\bibitem{Lencses:2024wib}
M.~Lencs\'es, A.~Miscioscia, G.~Mussardo and G.~Tak\'acs, \emph{{Ginzburg-Landau description for multicritical Yang-Lee models}}, \href{https://doi.org/10.1007/JHEP08(2024)224}{\emph{JHEP} {\bfseries 08} (2024) 224} [\href{https://arxiv.org/abs/2404.06100}{{\ttfamily 2404.06100}}].

\bibitem{Yang:1952be}
C.-N.~Yang and T.D.~Lee, \emph{{Statistical theory of equations of state and phase transitions. 1. Theory of condensation}}, \href{https://doi.org/10.1103/PhysRev.87.404}{\emph{Phys. Rev.} {\bfseries 87} (1952) 404}.

\bibitem{Lee:1952ig}
T.D.~Lee and C.-N.~Yang, \emph{{Statistical theory of equations of state and phase transitions. 2. Lattice gas and Ising model}}, \href{https://doi.org/10.1103/PhysRev.87.410}{\emph{Phys. Rev.} {\bfseries 87} (1952) 410}.

\bibitem{Fisher:1978pf}
M.E.~Fisher, \emph{{Yang-Lee Edge Singularity and $\varphi^3$ Field Theory}}, \href{https://doi.org/10.1103/PhysRevLett.40.1610}{\emph{Phys. Rev. Lett.} {\bfseries 40} (1978) 1610}.

\bibitem{Cardy:1985yy}
J.L.~Cardy, \emph{{Conformal Invariance and the Yang-lee Edge Singularity in Two-dimensions}}, \href{https://doi.org/10.1103/PhysRevLett.54.1354}{\emph{Phys. Rev. Lett.} {\bfseries 54} (1985) 1354}.

\bibitem{Cardy:2023lha}
J.~Cardy, \emph{{The Yang-Lee Edge Singularity and Related Problems}},  \href{https://arxiv.org/abs/2305.13288}{{\ttfamily 2305.13288}}.

\bibitem{Zamolodchikov:1989fp}
A.B.~Zamolodchikov, \emph{{Integrals of Motion and $S$-Matrix of the (Scaled) $T=T_c$ Ising Model with Magnetic Field}}, \href{https://doi.org/10.1142/S0217751X8900176X}{\emph{Int. J. Mod. Phys. A} {\bfseries 4} (1989) 4235}.

\bibitem{Zamolodchikov:1989hfa}
A.B.~Zamolodchikov, \emph{{Integrable field theory from conformal field theory}}, {\emph{Adv. Stud. Pure Math.} {\bfseries 19} (1989) 641}.

\bibitem{Wilson:1973jj}
K.G.~Wilson and J.B.~Kogut, \emph{{The Renormalization group and the $\epsilon$ expansion}}, \href{https://doi.org/10.1016/0370-1573(74)90023-4}{\emph{Phys. Rept.} {\bfseries 12} (1974) 75}.

\bibitem{Cardy:1989fw}
J.L.~Cardy and G.~Mussardo, \emph{{$S$ Matrix of the Yang-Lee Edge Singularity in Two-Dimensions}}, \href{https://doi.org/10.1016/0370-2693(89)90818-6}{\emph{Phys. Lett. B} {\bfseries 225} (1989) 275}.

\bibitem{Fonseca:2001dc}
P.~Fonseca and A.~Zamolodchikov, \emph{{Ising field theory in a magnetic field: Analytic properties of the free energy}}, \href{https://doi.org/10.1023/a:1022147532606}{\emph{Journal of Statistical Physics} {\bfseries 110} (2003) 527} [\href{https://arxiv.org/abs/hep-th/0112167}{{\ttfamily hep-th/0112167}}].

\bibitem{Zamolodchikov:1987ti}
A.B.~Zamolodchikov, \emph{{Renormalization Group and Perturbation Theory Near Fixed Points in Two-Dimensional Field Theory}}, {\emph{Sov. J. Nucl. Phys.} {\bfseries 46} (1987) 1090}.

\bibitem{Yurov:1989yu}
V.P.~Yurov and A.B.~Zamolodchikov, \emph{{Truncated Conformal Space Approach to Scale Lee-Yang Model}}, \href{https://doi.org/10.1142/S0217751X9000218X}{\emph{Int. J. Mod. Phys. A} {\bfseries 5} (1990) 3221}.

\bibitem{Yurov:1991my}
V.P.~Yurov and A.B.~Zamolodchikov, \emph{{Truncated fermionic space approach to the critical 2-D Ising model with magnetic field}}, \href{https://doi.org/10.1142/S0217751X91002161}{\emph{Int. J. Mod. Phys. A} {\bfseries 6} (1991) 4557}.

\bibitem{Kausch:1996vq}
H.~Kausch, G.~Takacs and G.~Watts, \emph{{On the relation between $\Phi_{(1,2)}$ and $\Phi_{(1,5)}$ perturbed minimal models}}, \href{https://doi.org/10.1016/S0550-3213(97)00056-4}{\emph{Nucl. Phys. B} {\bfseries 489} (1997) 557} [\href{https://arxiv.org/abs/hep-th/9605104}{{\ttfamily hep-th/9605104}}].

\bibitem{Cappelli:1986hf}
A.~Cappelli, C.~Itzykson and J.B.~Zuber, \emph{{Modular invariant partition functions in two dimensions}}, \href{https://doi.org/10.1016/0550-3213(87)90155-6}{\emph{Nucl. Phys. B} {\bfseries 280} (1987) 445}.

\bibitem{Quella:2006de}
T.~Quella, I.~Runkel and G.M.T.~Watts, \emph{{Reflection and transmission for conformal defects}}, \href{https://doi.org/10.1088/1126-6708/2007/04/095}{\emph{JHEP} {\bfseries 04} (2007) 095} [\href{https://arxiv.org/abs/hep-th/0611296}{{\ttfamily hep-th/0611296}}].

\bibitem{Ardonne_2011}
E.~Ardonne, J.~Gukelberger, A.W.W.~Ludwig, S.~Trebst and M.~Troyer, \emph{Microscopic models of interacting yang-lee anyons}, \href{https://doi.org/10.1088/1367-2630/13/4/045006}{\emph{New Journal of Physics} {\bfseries 13} (2011) 045006} [\href{https://arxiv.org/abs/1012.1080}{{\ttfamily 1012.1080}}].

\bibitem{EliasMiro:2022pua}
J.~Elias~Miro and J.~Ingoldby, \emph{{Effective Hamiltonians and Counterterms for Hamiltonian Truncation}}, \href{https://doi.org/10.1007/JHEP07(2023)052}{\emph{JHEP} {\bfseries 07} (2023) 052} [\href{https://arxiv.org/abs/2212.07266}{{\ttfamily 2212.07266}}].

\bibitem{Chang:2018iay}
C.-M.~Chang, Y.-H.~Lin, S.-H.~Shao, Y.~Wang and X.~Yin, \emph{{Topological Defect Lines and Renormalization Group Flows in Two Dimensions}}, \href{https://doi.org/10.1007/JHEP01(2019)026}{\emph{JHEP} {\bfseries 01} (2019) 026} [\href{https://arxiv.org/abs/1802.04445}{{\ttfamily 1802.04445}}].

\bibitem{Lassig:1991an}
M.~Lassig, \emph{{New hierarchies of multicriticality in two-dimensional field theory}}, \href{https://doi.org/10.1016/0370-2693(92)90581-N}{\emph{Phys. Lett. B} {\bfseries 278} (1992) 439}.

\bibitem{Ahn:1992qi}
C.-r.~Ahn, \emph{{RG flows of nonunitary minimal CFTs}}, \href{https://doi.org/10.1016/0370-2693(92)90683-U}{\emph{Phys. Lett. B} {\bfseries 294} (1992) 204} [\href{https://arxiv.org/abs/hep-th/9202028}{{\ttfamily hep-th/9202028}}].

\bibitem{Martins:1992ht}
M.J.~Martins, \emph{{Renormalization group trajectories from resonance factorized S-matrices}}, \href{https://doi.org/10.1103/PhysRevLett.69.2461}{\emph{Phys. Rev. Lett.} {\bfseries 69} (1992) 2461} [\href{https://arxiv.org/abs/hep-th/9205024}{{\ttfamily hep-th/9205024}}].

\bibitem{Martins:1992yk}
M.J.~Martins, \emph{{Exact resonance A-D-E S-matrices and their renormalization group trajectories}}, \href{https://doi.org/10.1016/0550-3213(93)90018-K}{\emph{Nucl. Phys. B} {\bfseries 394} (1993) 339} [\href{https://arxiv.org/abs/hep-th/9208011}{{\ttfamily hep-th/9208011}}].

\bibitem{Ravanini:1994pt}
F.~Ravanini, M.~Stanishkov and R.~Tateo, \emph{{Integrable perturbations of CFT with complex parameter: The $M(3/5)$ model and its generalizations}}, \href{https://doi.org/10.1142/S0217751X96000304}{\emph{Int. J. Mod. Phys. A} {\bfseries 11} (1996) 677} [\href{https://arxiv.org/abs/hep-th/9411085}{{\ttfamily hep-th/9411085}}].

\bibitem{Dorey:2000zb}
P.~Dorey, C.~Dunning and R.~Tateo, \emph{{New families of flows between two-dimensional conformal field theories}}, \href{https://doi.org/10.1016/S0550-3213(00)00185-1}{\emph{Nucl. Phys. B} {\bfseries 578} (2000) 699} [\href{https://arxiv.org/abs/hep-th/0001185}{{\ttfamily hep-th/0001185}}].

\bibitem{Fei:2014xta}
L.~Fei, S.~Giombi, I.R.~Klebanov and G.~Tarnopolsky, \emph{{Three loop analysis of the critical $O(N)$ models in $6-\epsilon$ dimensions}}, \href{https://doi.org/10.1103/PhysRevD.91.045011}{\emph{Phys. Rev. D} {\bfseries 91} (2015) 045011} [\href{https://arxiv.org/abs/1411.1099}{{\ttfamily 1411.1099}}].

\bibitem{Zamolodchikov:1986gt}
A.B.~Zamolodchikov, \emph{{Irreversibility of the Flux of the Renormalization Group in a 2D Field Theory}}, {\emph{JETP Lett.} {\bfseries 43} (1986) 730}.

\bibitem{Castro-Alvaredo:2017udm}
O.A.~Castro-Alvaredo, B.~Doyon and F.~Ravanini, \emph{{Irreversibility of the renormalization group flow in non-unitary quantum field theory}}, \href{https://doi.org/10.1088/1751-8121/aa8a10}{\emph{J. Phys. A} {\bfseries 50} (2017) 424002} [\href{https://arxiv.org/abs/1706.01871}{{\ttfamily 1706.01871}}].

\bibitem{Bender:2018pbv}
C.M.~Bender, N.~Hassanpour, S.P.~Klevansky and S.~Sarkar, \emph{{$\mathcal{PT}$-symmetric quantum field theory in $D$ dimensions}}, \href{https://doi.org/10.1103/PhysRevD.98.125003}{\emph{Phys. Rev. D} {\bfseries 98} (2018) 125003} [\href{https://arxiv.org/abs/1810.12479}{{\ttfamily 1810.12479}}].

\bibitem{Dotsenko:1984nm}
V.S.~Dotsenko and V.A.~Fateev, \emph{{Conformal Algebra and Multipoint Correlation Functions in Two-Dimensional Statistical Models}}, \href{https://doi.org/10.1016/0550-3213(84)90269-4}{\emph{Nucl. Phys. B} {\bfseries 240} (1984) 312}.

\bibitem{Dotsenko:1984ad}
V.S.~Dotsenko and V.A.~Fateev, \emph{{Four Point Correlation Functions and the Operator Algebra in the Two-Dimensional Conformal Invariant Theories with the Central Charge $c \leq 1$}}, \href{https://doi.org/10.1016/S0550-3213(85)80004-3}{\emph{Nucl. Phys. B} {\bfseries 251} (1985) 691}.

\bibitem{DiFrancesco:1997nk}
P.~Di~Francesco, P.~Mathieu and D.~Senechal, \emph{{Conformal Field Theory}}, Graduate Texts in Contemporary Physics, Springer-Verlag, New York (1997), \href{https://doi.org/10.1007/978-1-4612-2256-9}{10.1007/978-1-4612-2256-9}.

\bibitem{Smirnov:1990vm}
F.A.~Smirnov, \emph{{Reductions of the sine-Gordon model as a perturbation of minimal models of conformal field theory}}, \href{https://doi.org/10.1016/0550-3213(90)90255-C}{\emph{Nucl. Phys. B} {\bfseries 337} (1990) 156}.

\bibitem{Freund:1989jq}
P.G.O.~Freund, T.R.~Klassen and E.~Melzer, \emph{{S-Matrices for Perturbations of Certain Conformal Field Theories}}, \href{https://doi.org/10.1016/0370-2693(89)91165-9}{\emph{Phys. Lett. B} {\bfseries 229} (1989) 243}.

\bibitem{Koubek:1991dt}
A.~Koubek and G.~Mussardo, \emph{{$\phi_{1,2}$ deformation of the $M_{2,2n+1}$ conformal minimal models}}, \href{https://doi.org/10.1016/0370-2693(91)91052-W}{\emph{Phys. Lett. B} {\bfseries 266} (1991) 363}.

\bibitem{Smirnov:1991uw}
F.A.~Smirnov, \emph{{Exact $S$-matrices for $\phi_{1,2}$-perturbated minimal models of conformal field theory}}, \href{https://doi.org/10.1142/S0217751X91000745}{\emph{Int. J. Mod. Phys. A} {\bfseries 6} (1991) 1407}.

\bibitem{Zamolodchikov:1989cf}
A.B.~Zamolodchikov, \emph{{Thermodynamic Bethe Ansatz in Relativistic Models. Scaling Three State Potts and Lee-yang Models}}, \href{https://doi.org/10.1016/0550-3213(90)90333-9}{\emph{Nucl. Phys. B} {\bfseries 342} (1990) 695}.

\bibitem{Zamolodchikov:1995xk}
A.B.~Zamolodchikov, \emph{{Mass scale in the sine-Gordon model and its reductions}}, \href{https://doi.org/10.1142/S0217751X9500053X}{\emph{Int. J. Mod. Phys. A} {\bfseries 10} (1995) 1125}.

\bibitem{Destri:1990ps}
C.~Destri and H.J.~de~Vega, \emph{{New exact results in affine Toda field theories: Free energy and wave function renormalizations}}, \href{https://doi.org/10.1016/0550-3213(91)90540-E}{\emph{Nucl. Phys. B} {\bfseries 358} (1991) 251}.

\bibitem{Leitner:2018iyf}
M.~Leitner, \emph{{The $(2,5)$ minimal model on genus two surfaces}},  \href{https://arxiv.org/abs/1801.08387}{{\ttfamily 1801.08387}}.

\bibitem{Fitzpatrick:2023aqm}
A.L.~Fitzpatrick, E.~Katz and Y.~Xin, \emph{{Lightcone Hamiltonian for Ising Field Theory I: $T<T_c$}},  \href{https://arxiv.org/abs/2311.16290}{{\ttfamily 2311.16290}}.

\bibitem{Zamolodchikov:2001dz}
A.~Zamolodchikov, \emph{{Scaling Lee-Yang model on a sphere. 1. Partition function}}, \href{https://doi.org/10.1088/1126-6708/2002/07/029}{\emph{JHEP} {\bfseries 07} (2002) 029} [\href{https://arxiv.org/abs/hep-th/0109078}{{\ttfamily hep-th/0109078}}].

\end{thebibliography}\endgroup

\end{document}